\documentclass[aps,prl,twocolumn,letterpaper,superscriptaddress,showpacs,floatfix, 10pt,nofootinbib]{revtex4-2}
\usepackage{graphicx}
\usepackage{CJK}
\usepackage{tikz}
\usepackage{svg}
\usepackage{verbatim}
\usepackage{lineno}
\usepackage{bm}
\usepackage{times}
\usepackage{subfigure}
\usepackage{amsmath}
\usepackage{physics}
\usepackage{subfiles}
\usepackage{natbib}
\usepackage{lipsum}
\usepackage[normalem]{ulem}
\makeatletter

\usepackage{titlesec}

\titleformat*{\section}{\large\centering\bfseries}

\usepackage{xcolor}

\begin{document}

\begin{CJK*}{UTF8}{bsmi}

\title{Electronic Excitations in the Bulk of Fractional Quantum Hall States}

\author{Xinlei Yue}
\affiliation{Department of Condensed Matter Physics, Weizmann Institute of Science, Rehovot, Israel 7610001}
\author{Ady Stern}
\affiliation{Department of Condensed Matter Physics, Weizmann Institute of Science, Rehovot, Israel 7610001}
\begin{abstract}
We analyze electronic excitations (excitations generated by adding or removing one electron) in the bulk of fractional quantum Hall states in Jain sequence states, using composite fermion Chern-Simons field theory. 
Starting from meanfield approximation in which gauge field fluctuations are neglected, we use symmetry to constrain the possible composite fermion states contributing to electronic Green's function and expect discrete infinitely-sharp peaks in the electronic spectral function. We further consider the electronic excitations in particle-hole conjugate fractional quantum hall states. Gauge field fluctuations play an increasingly important role in the electron spectral function as the filling factor approaches $1/2$, and  evolve the discrete coherent peaks into a broad continuum even in the absence of impurities. At that limit, we calculate the dressed electron spectral function via linked cluster approximation from the low to intermediate energy range. Finally, we compare our results with recent experiments. 
\end{abstract}
\maketitle
\end{CJK*}
Fractional quantum Hall(FQH)\cite{Tsui1982,laughlin1983,english1987, MOORE1991362,halperin2020fractional} states are famous for hosting anyonic excitations that are neither fermionic nor bosonic. 
 The anyonic excitations belong to the Hilbert space $\mathcal{H}_{N}$ of a fixed number of electrons $N$. 
 %$\mathcal{H}_{N}$, spanned by \textit{\textbf{fixed}} $N$ electrons.
The term "anyon" implies that these are localized excitations of a particle-like nature that accumulate a fractional geometrical phase upon braiding.  
%\ASc{Why do we need the following sentences?} Start from a many-body state $\ket{\vec{\mathbf{r}}_1,\vec{\mathbf{r}}_2}_N$ with two such anyons locating far from each other. If we adiabatically interchange their position, we arrive at a many-body state $\mathcal{B}\ket{\vec{\mathbf{r}}_1,\vec{\mathbf{r}}_2}_N$ differs from the starting state by a complex phase factor, $e^{i\phi}\ket{\vec{\mathbf{r}}_1,\vec{\mathbf{r}}_2}_N$, or even a matrix, $B\ket{\vec{\mathbf{r}}_1,\vec{\mathbf{r}}_2}_N$. The underlying quantum states are therefore characterized as Abelian or non-Abelian topological ordered depending on the properties of $\mathcal{B}$.

However, single anyons cannot be added or removed to an FQH system. Rather, only full electrons may be exchanged by the system with the outside world. 
%There are yet other sets of excitations, electronic excitations, for example, that differ from the anyonic excitations described above. 
The electronic excitations describe adding(removing) an electron to(from) the system and therefore correspond to many-body states living in the Hilbert space, $\mathcal{H}_{N\pm 1}$, spanned by $N\pm 1$ electrons. Experiments\cite{PhysRevLett.69.3804, EISENSTEIN1994393} in the early days revealed the properties of these electronic excitations via tunneling between weakly coupled double-layer heterostructures. In a strong magnetic field these experiments observed tunneling spectra with features that are almost independent of the filling factors. Primary among these features is a highly suppressed tunneling conductance at low voltage. Theory attempts\cite{PhysRevB.48.14694, Halperin1993, PhysRevLett.71.1435,PhysRevB.50.8078,levitov1997semiclassical} successfully understand the low voltage \textbf{\textit{incoherent}} tunneling characteristics as originating from Coulomb interaction and weak dissipation. Distinct from this incoherent broad continuum, Jain\cite{PhysRevLett.94.186808} proposes the existence of \textit{\textbf{coherent}} electronic excitations, which do depend on the underlying topological order and contribute sharp peaks in tunneling microscopes. More recently, scanning tunneling microscope experiments\cite{farahi2023broken, hu2023highresolution} have been performed on FQH states in graphene-based systems, and interest in studying the electronic excitation in the bulk of FQH states has been invoked. Interestingly, coherent sharp peaks and incoherent continuums are found at different fillings, which raises the question of the relation between these two different phenomena.
%However, the difference/relation between coherent and incoherent excitation is unclear. More generally, discussion on what we should expect in the electronic excitation spectra of different FQH states is absent.

In this paper, we use composite fermion (CF) theory to analyze the relationship between sharp peaks and incoherent continuums observed in tunneling experiments. 
%answer the above questions using composite fermion (CF) theory. 
Composite fermion theory maps the problem of electrons in an FQH state to a problem of composite fermions in an Integer Quantum Hall state, which interacts with one another through a dynamical Chern-Simons gauge field. 
%In particular, we use CF theory to describe the many-body states in low-energy Hilbert space, dramatically reducing the analysis's complexity. 
We start with meanfield approximation to this mapping, which neglects the Chern-Simons(CS) gauge field fluctuations. Under this approximation, we describe the many-body states after adding or removing an electron and use symmetry to limit the resulting CF states that are allowed. At this level, due to the gapped structure of the CF states, we expect discrete infinitely sharp peaks in the one-electron spectral function for all the FQH states in Jain's sequence.
We consider the electronic excitations in the Jain series and its P-H conjugate state and highlight a subtlety in using composite fermion theory for the latter.

Following mean field analysis, we proceed to study the effect of gauge field fluctuations. Utilizing the fact that the tunneling electron is hard to dissipate due to low longitudinal conductance. We adapt the linked cluster approximation to analyze the fate of the meanfield CF states we found once fluctuations are taken into account. We show that the electronic spectral functions evolve from coherent discrete peaks to the incoherent continuum as the filling factor approaches 1/2, even in the absence of impurities. The coherent and incoherent electronic excitations are, therefore, unified in this picture and are shown to evolve from one to the other by gauge field fluctuations. Near half filling,  the electron operators do not participate in the low energy dynamics of the background electrons, which is described by composite Fermi liquid (CFL)\cite{PhysRevB.47.7312}. In this limit, we study the effect of density fluctuation and calculate the dressed electron spectral function using linked cluster approximation. We find an analytical expression for the shape of the spectral function, which contains a sharp edge on the low-energy side and decays algebraically on the high-energy side. We then explain how Landau quantization of CFL will evolve this broad continuum into coherent sharp peaks as the filling factor significantly deviates from half-filling. 
Finally, we compare our results with recent tunneling microscope experiments.

We describe the electronic excitations by the standard one-electron spectral function $A(\lambda,\omega)=-2\Im G_R(\lambda,\omega)$ where $\lambda$ denotes the set of quantum numbers of the tunneling electron.
%chosen to diagonalize the retarded electron Green's function $G_R$. 
%In our case, particle(hole) spectral function $A_{+}(\omega)$\Big($A_{-}(\omega)$\Big) is easier to obtain. 
In the zero temperature limit,

\begin{equation}
    \begin{aligned}
    \label{spectal}
        A(\lambda,\omega)&=A_{+}(\lambda,\omega)+A_{-}(\lambda,\omega)\\
        A_{+}(\lambda,\omega)&=2\pi \sum_\alpha \delta(\omega - E_{\alpha}+E_{g}) O_{\alpha;\lambda}^+\\
        A_{-}(\lambda,\omega)&=2\pi \sum_\beta \delta(\omega +E_{\beta}-E_{g}) O_{\beta;\lambda}^-,
    \end{aligned}
\end{equation}
%$\mp\mu_\pm=\Theta(N\pm1)-\Theta(N)$, $\Theta(n)$ is the environment-related energy which is a function of electron number and gets minimized when there's $N$ electron in the system
with 
\begin{equation}
\begin{aligned}
    O_{\alpha;\lambda}^+&=\Bigl|\, _{N+1}\!\!\bra{\alpha}c_{\lambda}^\dagger \ket{g}_N\Bigr|^2\\
    O_{\beta;\lambda}^-&=\Bigl|\, _{N-1}\!\!\bra{\beta}c_{\lambda} \ket{g}_N\Bigr|^2\\
\end{aligned}
\end{equation}
where $\ket{g}_N$ is the ground state in $\mathcal{H}_{N-}$, $\alpha$($\beta$) labels the many-electron eigenstates with eigenenergy $E_{\alpha}$($E_{\beta}$) in $\mathcal{H}_{N+ 1}$($\mathcal{H}_{N- 1}$), $c_\lambda$($c^\dagger_\lambda$) is the electronic operator that destroys (creates) an electron with quantum number $\lambda$. The many-body states, resulting from strong interactions, are highly entangled and, therefore, very hard to describe using electron language. Fortunately, we could employ CF theory to make progress.

\begin{figure}
    \centering
    \includegraphics[width = 240pt]{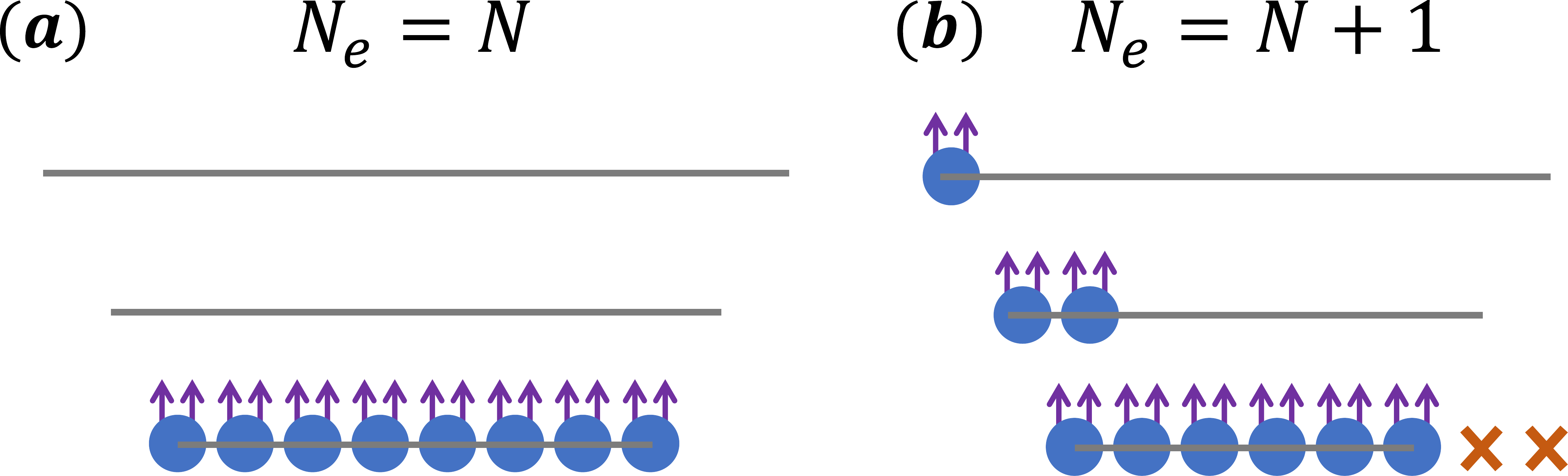}
    \caption{Examples of CF states of many-electron states. The horizontal lines correspond to $\Lambda$-levels. From top to bottom, the lines corresponding to $\Lambda$-levels with lower index and their length indicate the corresponding degeneracy. We use a solid circle with two arrows representing the CF particle and an empty circle for the CF hole. The single CF states have lower and lower z-angular momentum from left to right. (a) CF state corresponds to the ground state of $\nu=1/3$. (b) lowest energy CF state that could have nonzero overlap with $c^\dagger_{QQQ}\ket{g}$, where $c_{QQQ}^\dagger$ add an electron(in the lowest Landau level) to the north pole of the sphere and $Q$ is the monopole strength in the sphere. The red crosses emphasize that the degeneracy on each $\Lambda$-level decreases because of the added electron.}
    %(c) lowest energy CF state that could have nonzero overlap with $c_{QQQ}\ket{g}$, the green "plus" signs emphasize that the degeneracy on each $\Lambda$-level increase due to the removed electron.}
    \label{CF_config}
\end{figure}

\begin{figure}
    \centering
    \includegraphics[width = 240pt]{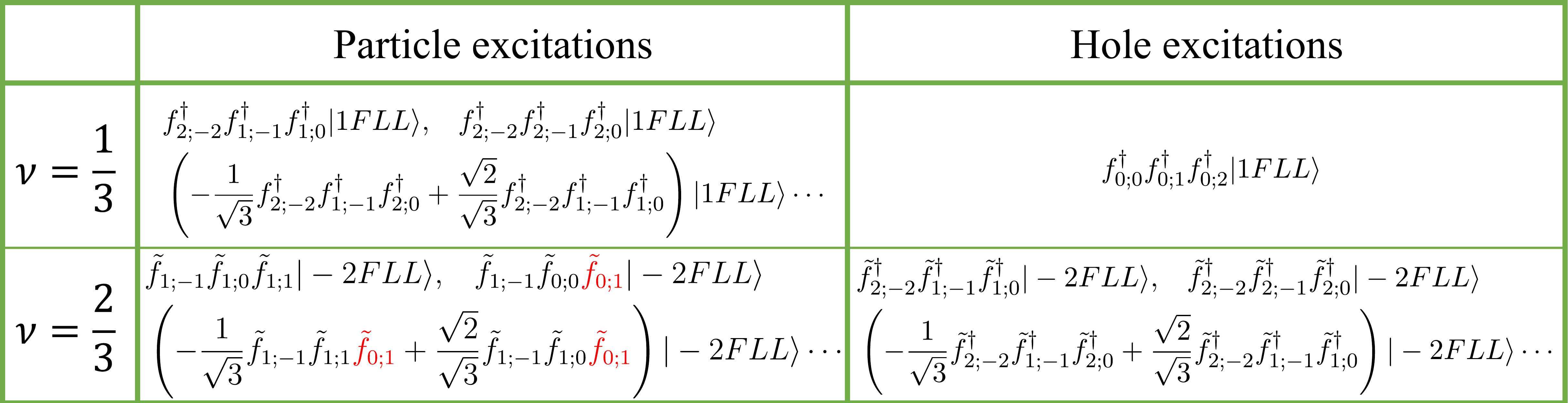}
    \caption{Low energy CF states corresponding to particle(hole) electronic excitations $c^\dagger_{QQQ}\ket{g}$
    ($c_{QQQ}\ket{g}$). $f_{i;j}^\dagger$($\tilde{f}_{i;j}^\dagger$) is the creation operator of CF in the $i$th $\Lambda$-levels with the $i+j-1$th largest(smallest) $z$-angular momentum. $|nFLL\rangle$ shows the ground state of  CFs at specific fillings with $n$ filled $\Lambda$-levels. The minus sign of $n$ emphasizes the CFs feel an effective magnetic field opposite to the external magnetic field.
    For example, the CF state in Fig.\ref{CF_config}(b) is denoted by  $f_{2;-2}^\dagger f_{1;-1}^\dagger f_{1;0}^\dagger|1FLL\rangle$. The red operators in some CF states indicate that these CF states involve CF excitations in the lowest $\Lambda$-level for $1/2<\nu<1$. The existence of such states \textbf{\textit{break}} the particle-hole symmetry between the particle-hole conjugate FQH states. (The coefficients in superpositioned CF states are obtained in large monopole strength limit.)}
    \label{CF_state}
\end{figure}
In CF theory, Chern-Simons transformation attaches an even number of flux quanta, say $2\Tilde{\phi}$, to each electron, turning it into a composite fermion. 
After the transformation, the original strongly interacting electrons at density $n$, magnetic field $B$ and filling factor $\nu=p/(2p+1)$ (with $p$ an integer) are described by CFs at the same density, reduced magnetic field $\Delta B=B-2\Tilde{\phi}\Phi_0n$ and filling factor $p$, coupled to electromagnetic and Chern-Simons gauge fields. Progress is made by first considering meanfield approximation in which density fluctuations are neglected. Importantly, even at this level, the theory correctly identifies the difference between adding (or removing) an electron {\it vs.} a composite fermion to a gapped state. In particular, when an electron is added to a $p/(2p+1)$ state, it reduces $\Delta B$ by two flux quanta. Consequently, each of the $p$ filled levels loses two states, and thus, a total of $2p+1$ composite fermions are added to the otherwise empty Landau levels, see Fig.\ref{CF_config}. The single CF states that these composite fermions may occupy must be chosen such that the constraints imposed by the quantum numbers are respected. 

%The system is then described by free CF in an effective magnetic field $B^*$. Further correction could be made by taking into account the gauge field fluctuation perturbatively or attaching Jastrow factors to get guessed wave functions. CF theory's great success in describing the FQH states implies that CF theory, even at the meanfield level, captures the key ingredients of low energy Hilbert space. 

These constraints are handled in Ref. \cite{PhysRevLett.94.186808} by considering an FQH system ($\nu=p/(2p+1)$ with positive $p$) on a sphere and finding the meanfield CF states with the lowest energy that have a significant contribution to the spectral function. For the particle (electron) excitation, the lowest energy CF states contain $p$ and $p+1$ CF holes in the $p+1$th and $p$th $\Lambda$-levels (the CF Landau levels). While for the hole excitation, the lowest energy CF states contain $p+2$ and $p-1$ CF holes in the $p-1$th and $p-2$th $\Lambda$-levels.
For completeness, we review this analysis in Appendix A. A similar analysis may be carried out on a planar geometry.

%\ASc{What do you mean "we approximate"?}\XYc{I refined the statement}
{ Here, we proceed with spherical geometry and consider electron tunneling from/to the north pole of the sphere. In the large magnetic field limit, we could just focus on the lowest Landau level. At low energy (relative to the cyclotron energy), this tunneling electron is characterized by $\lambda=QQQ$ where the three numbers denote the monopole strength, the Landau level index, and the azimuthal angular momentum. In later discussion, we will focus on analysing the spectral function $A(QQQ,\omega)$ and omit the label ($\lambda=QQQ$) from now until necessary. }In Fig.\ref{CF_state}, we show the possible low energy CF states for small $p$ constrained by full rotational symmetry. 
These CF states will contribute delta-function-like peaks in the spectral function according to,
\begin{equation}
    A_+(\omega)=\sum_\alpha O_\alpha^+ A_\alpha (\omega),
\end{equation}
at this meanfield level\footnote{It's natural to expect the discrete peaks predicted at the meanfield level to merge into a broad continuum because of impurity scattering. However, we will see that fluctuations will merge these discrete peaks into a broad continuum as the filling factor approaches $1/2$ even in the absence of impurity. }. Although we could not calculate the corresponding spectral weight of these CF states explicitly in this approach, a description of the envelope-line of these CF spectral functions is possible based on the analysis of $\nu=1/2$. We defer this discussion to the later part of the paper.

%Although we cannot calculate their corresponding spectral weight in this approach, we expect the spectral weight corresponding to higher energy CF states to be smaller because higher energy CF states have more spread density and contain more higher Landau level components. Consequently, on both particle and hole sides, we expect the peak at the lowest activation energy to have the highest spectral weight. 

Applying mean field CF theory to the analysis of FQHE states with filling factor $1/2<\nu<1$ requires some care. The Jain series for that range satisfies $\nu=\frac{p+1}{2p+1}$. It may be viewed as a hole Jain series state of $\nu_h=\frac{p}{2p+1}$ in a vacuum of $\nu=1$. As such, adding or removing electrons from the system corresponds to removing or adding holes from the $\nu_h=p/(2p+1)$ state. Importanttly, there is no separate process in which a bulk electron is added or removed from the $\nu=1$ state. In the $\nu=(p+1)/(2p+1)$  state, the number of electrons in the $\nu=1$ vacuum may be changed only by the variation of a magnetic field, which changes the number of states in the $\nu=1$ Landau level. When considered this way, it becomes apparent that the electronic addition and removal spectra are interchanged between $\nu=p/(2p+1)$ and $\nu=(p+1)/(2p+1)$, as a consequence of the two states being related by particle-hole symmetry. 

This particle-hole symmetry is somewhat subtle to realize when the $\nu=(p+1)/(2p+1)$ is analyzed by means of flux attachment of $2\Phi_0$ to each electron. When that is done, the composite fermion state is a $\nu_{CF}=-p-1$. Seemingly, the particle-hole symmetry is lost, since $p+1$ $\lambda$-levels are occupied, rather than $p$ levels for $\nu=p/(2p+1)$. However, more careful analysis will resolve this discrepancy. It is best illustrated using the formalism of $K$-matrices. For simplicity, we focus on the $p=1$ case, namely $\nu=2/3$. The $\nu=2/3$ state is described by
\begin{eqnarray}
    K=\left ( \begin{matrix} 1 & 0 \\ 0 & -3 \end{matrix} \right )
\label{Kmatrix}
\end{eqnarray}
which includes a fully filled electron Landau level and $1/3$-filled Landau level of holes on top of it. Again, when an additional electron is added to the bulk, it can only be added to the $1/3$ hole state. Therefore, the added electron only couples to the effective CS gauge field corresponding to filling $1/3$ hole state via $-3a_{2,0}\delta(\mathbf{x}-\mathbf{x_0})$, where $a_{I,\mu}$ labels the $\hat{x}_\mu$ component of $I\,$th CS field.

The $K$-matrix (\ref{Kmatrix}) may be transformed by an $SL(Z,2)$ transformation $\left (\begin{matrix} 1 & 0 \\ 2 & 1 \end{matrix}\right )$ to 
\begin{eqnarray}
    \widetilde{K}=\left ( \begin{matrix} 1 & 2 \\ 2 & 1 \end{matrix} \right )
\label{K1matrix}
\end{eqnarray}
indicating that the two matrices (\ref{Kmatrix}) and (\ref{K1matrix}) describe the same topological order. The $K$-matrix (\ref{K1matrix}) describes the flux attachment procedure, as may be understood by noticing that $\widetilde{K}=\left ( \begin{matrix} -1 & 0 \\ 0 & -1 \end{matrix} \right )+\left ( \begin{matrix}2 & 2 \\ 2 & 2 \end{matrix} \right )$. The previous coupling between the inserted electron and CS field is transformed to $-3\widetilde{a}_{2,0}\delta(\mathbf{x}-\mathbf{x_0})$, which involves just one filled $\Lambda$-level. Thus, one of the two $\Lambda$-levels is frozen, and particle-hole symmetry with the $\nu=1/3$ electric state is kept. In general, for filling factors $\nu=(p+1)/(2p+1)$, to recover the particle-hole symmetry, one just effectively "freezes" the lowest $\Lambda$-level and analyzes the possible CF states via typical flux attachment method as $\nu=p/(2p+1)$ cases.

%  ==================
 
% It's interesting to consider the electronic excitation in the P-H conjugate states, $\nu=p/(2p+1)$ and $\nu=(p+1)/(2p+1)>1/2$. We could predict the electronic excitation spectrum in the same approach, as summarized in Table.1. The only difference is that we get an excess of $2\Tilde{\phi}p+1$ CF holes(particles) when we add an electron(hole) $c_{QQQ}^\dagger$ to the system. 

\begin{figure}
    \centering
    \includegraphics[width = 180pt]{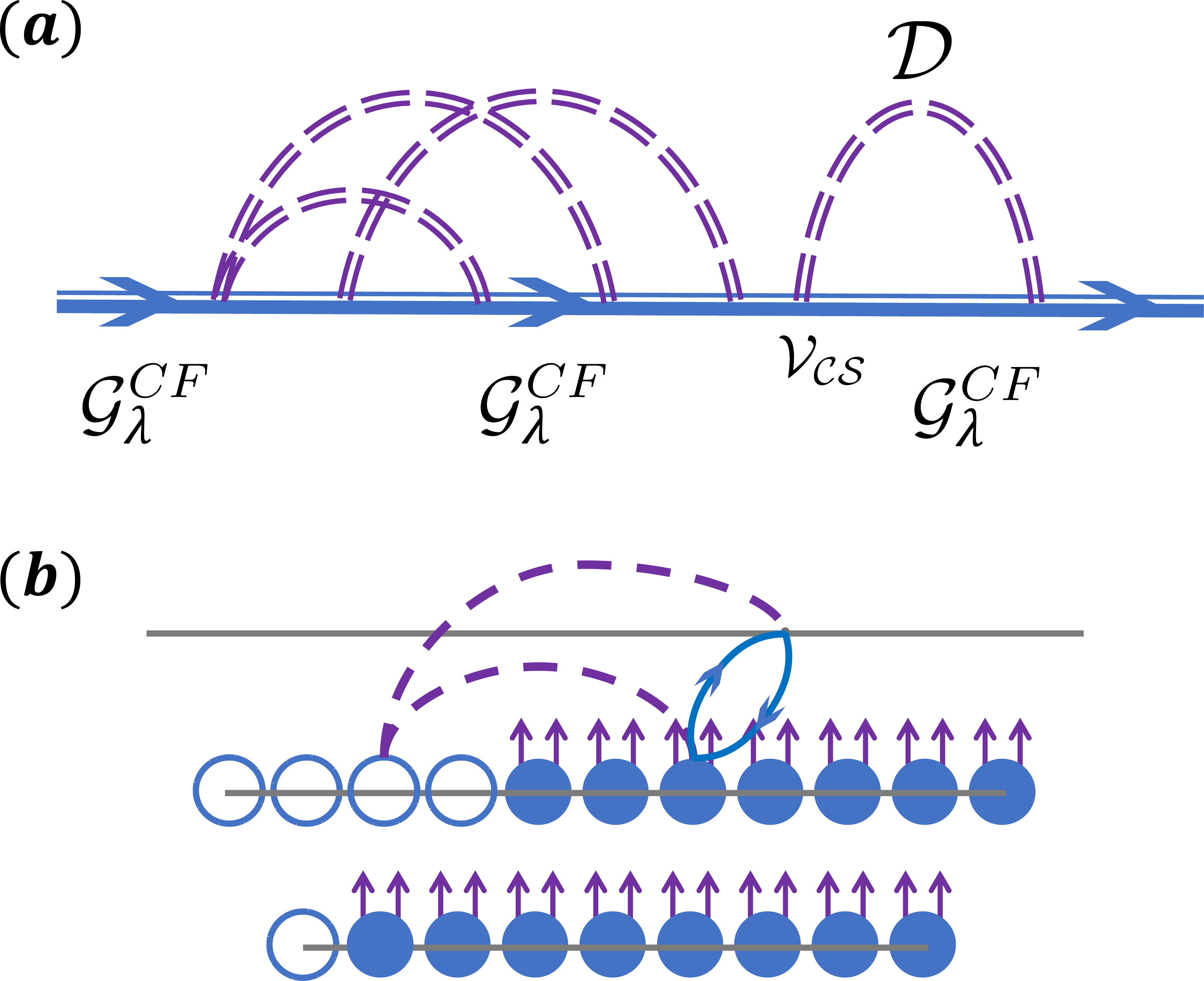}
    \caption{Illustration of scattering process appears in linked cluster approximation. (a) Example of linked cluster diagrams. Purple double dash lines correspond to the fully dressed CS propagator. Blue double lines represent propagators of CF states. In linked cluster approximation, it's assumed the unperturbed state could not be scattered off. Notice that in linked cluster approximation, we don't use the Gell-Mann and Low theorem, and not only connected graphs need to be considered. (b) An illustration of the scattering vertex appears in linked cluster approximation. It's clear from the figure that there is no energy-conserved process here.}
    \label{link_cluster_diagram}
\end{figure}

We now address the effect of gauge field fluctuations on the infinitely sharp peaks in the spectral function predicted by mean field theory. We expect two effects of the fluctuations - a shift of the peaks' energies and a change in the shape of the delta function peaks. Since mean field theory fails to predict the correct energy scale for composite fermion excitations(at least for small $p$), we focus on the spectral function's shape. 

%, we must consider the effect of fluctuations, which was neglected at the meanfield level. 
% To analyze the effects of CS field fluctuation, we artificially separate the scattering process into two classes: 1. scattering to CF states with similar or higher energy and 2. scattering to CF states with lower energy. Note that the CS field is a gapped mode; the first class of scattering is energy-unconserved virtual scattering, and the second class might involve an energy-conserved process.
% Basic quantum mechanics then tells us that the first-class scattering will redistribute the spectral weight, and the second-class scattering might result in a finite lifetime.

We analyze the effect of the fluctuations on the spectral function using 
%. Among such scattering processes, the dominant contribution will come from the scattering to CF states with the same energy. We thus adapted 
linked cluster approximation\cite{mahan2000, Halperin1993, PhysRevB.73.073306} in which the tunneling electron interacts with background electrons but is assumed to stay at its initial state. 
We expect this approximation to be valid here\footnote{Possible scattering to meanfield state with same energy is expected only to change the coupling vertex quantitatively.} because the Hall states have a very low longitudinal conductance and therefore the tunneled electron, as a point charge in a homogeneous bulk, is very hard to dissolve into the background.

Under this approximation, the dressed CF state Green's function is given by
\begin{equation}
    \mathcal{G}_\alpha^{LC} (t)=e^{F_2(t)+F_3(t)}\mathcal{G}_\alpha^{CF} (t).
    \label{lcexpression}
\end{equation}
where $\mathcal{G}_\alpha^{CF}=\Theta(t)e^{-i E_\alpha t }$ is the meanfield time-ordered Green's function for the operator $\hat{M_\alpha}$ that acts on the ground state $\ket{g}$ and results meanfield CF state $\hat{M_\alpha}\ket{g}=\ket{\alpha}$, and $E_\alpha$ is the meanfield excitation energy for the CF state $\ket{\alpha}$. The detailed expression for $F_2(t)$ and $F_3(t)$ are given in Appendix B (the subscript denotes the number of coupling vertices in the diagram). 
%\ASc{It looks strange that we start from $F_2$ and there is no $F_1$. Why do we do that?} 
In practice, the contribution mainly comes from
\begin{equation}
    F_2^1 (t)=-\int_0^{\infty}\dd u f(u)\left(1-e^{-iut}-iut\right)
\end{equation}
with\footnote{since the fluctuation effect is expected to be independent of particular geometry, we will use planar geometry hereinafter for simplicity}
\begin{equation}
    f(u)=-\frac{1}{ u^2}\! \!\!\int \!\!\! \frac{q \dd q}{\left(2\pi\right)^2} V_q^{\mu}V_{-q}^{\nu}  \Im[D_{\mu\nu}^r(q,u)]
\end{equation}
where $D_{\mu\nu}^r(q,u)$ is the retarded CS gauge field propagator at wave vector $q$ and frequency $u$, and $V_q^{\mu}$ is the coupling of the CF state with CS gauge field, whose detailed form is given in Appendix D.

%Since the scattering process considered in the linked cluster approximation is not energy conserved, 
One feature of linked cluster approximation is that the lowest energy peak remains infinitely sharp but transfers its weight to higher energy features, as one can see from the expansion
\begin{equation}
\begin{aligned}
    A^{LC}_\alpha (\omega+\omega_{min})=&e^{-\int_0^{\infty}\dd u f(u)}\\
    &\times\left(A_\alpha (\omega)+\!\!\!\int_{0}^{\infty}\!\!\!\! \dd u \,A_\alpha (\omega-u)f(u)+\dots \right),
\end{aligned}
    \label{expression_spec}
\end{equation}
where ellipses represent higher-order convolution terms. We find the weight of the lowest energy peak is suppressed by a factor $S=\exp(-\int_0^{\infty}d u f(u))$, and the energy is pushed down by an amount of $\omega_{min}=\int_0^\infty \dd u f(u)u$ as a consequence of level repulsion.

We evaluate the dressed spectral function of a \textit{single} peak predicted by meanfield CF theory, say $A_\alpha$ corresponding to meanfield CF state $\ket{\alpha}$. We will focus here on the large $p$ limit, which is shown below to be directly related to the transition of the electron spectral function from the sharp peaks to a broad continuum. The same approximation could be applied to small $p$ cases, and we present the numerical results in Appendix C. To do this, we approximate the CS gauge field propagator using its form at $\nu=1/2$ accompanied by the cutoffs of the integral governed by the filling factor. Since $F_2(t)$ will be dominated by the infrared part, we expect the contribution to be dominated by $D_{11}(q,\omega)$, which is more divergent at small energy and momentum, 
\begin{equation}
    D_{11}(q, \omega) \approx\left[i \frac{k_F}{2 \pi} \frac{\omega}{q}-q \ell_B \frac{E_C}{8 \pi}\right]^{-1}.
\end{equation}
Information about the specific filling factor hides in the cutoffs of the integral. For example, the suppression factor reads,%\ASc{You miss a minus sign there, no?}
\begin{equation}
S(p) =\exp \left(\int_{\bar{\Delta}}^{\infty} \mathrm{d} u \frac{1}{u^2} \int_{\frac{1}{R_{\Delta}}}^{\frac{1}{\ell_{\Delta}}} \frac{q \mathrm{~d} q}{(2 \pi)^2} V_q^1 V_{-q}^1 \operatorname{Im}\left[D_{11}^r(q, u)\right]\right)
\end{equation}
where we set the low energy cutoff to be the neutral excitation gap $\Tilde{\Delta}$ which was evaluated to be $\frac{E_C \pi}{2(2 p+1)} \frac{1}{\ln (2 p+1)}$ under the same scheme\cite{PhysRevB.52.5890} and the momentum cutoff is given by the form of the coupling $V_q^1$ (details of this analysis are given in Appendix E). Here $E_C=e^2/\epsilon\ell_B$, with $e$ the electron charge, $\epsilon$ the dielectric constant, and $\ell_B$ the magnetic length. We conclude that the spectral weight of the lowest energy peak vanishes in the large $p$ limit as 
\begin{equation}
    S(p)=\exp \left(-\frac{8(2 p+1) \ln ^2(2 p+1) {\omega_c^*}^2}{c_v\pi^5 E_c^2}\right).
    \end{equation}
    where $c_v\approx 2$ is a crossover value at which the coupling vertex changes its behavior and gives an effective cutoff scale (see Appendix E for details). More importantly, most of the spectral weight is transferred to higher energy. It may be calculated by Fourier transforming the following time-dependent function
    %and described by the dressed spectral function \ASc{How is a spectral weight described by a function of time, and not of energy? You mean to say that the spectral function is the FT of this?}.
\begin{equation}
    \begin{aligned}
        A_\alpha^{LC}(t)&=\exp(-i E_\alpha t + F_2(t))\\
        &\approx\exp(-i E_\alpha t-\frac{t^2{\omega_c^*}^2}{c_v\pi^3(2p+1)}\ln(2p+1) )
    \end{aligned}
\end{equation}
which results a Gaussian in frequency domain with $\text{FWHM}\propto \omega_c^*\sqrt{\ln{p}/p}$ at the meanfield energy $E_\alpha$ of the corresponding CF state. $\gamma$ is the Euler's constant, $\omega_c^*=eB/m^*$ with $e$ the electron charge, $B$ the applied magnetic field, and $m^*$ the effective mass of CFs at meanfield level. since the distance between consecutive meanfield peaks in the electron spectral function is expected to be of the order of the neutral excitation gap $\Tilde{\Delta}$, whose decay to zero at large $p$ is faster, we expect that even in the \textit{clean limit}, as the filling factor approaches $1/2$, the discrete peaks at small $p$ will evolve into a broad (incoherent) continuum due to gauge field fluctuations. If we take $\omega_c^*\approx 0.3 E_C$ obtained from exact diagonalization at $\nu=1/3$, we find the crossover value, $\text{FWHM}\sim \Tilde{\Delta}$, to be $p_c\approx 9$

Switching to large $p$ limit is instructive. In this limit, the system is compressible, and it becomes important to discuss how the density fluctuation dresses the electron spectral function. We use Halperin-Lee-Read theory\cite{PhysRevB.47.7312} as the effective low energy theory for the large $p$ limit and explore the results.
Notice that the large magnetic field suppresses the exchange interaction. The tunneling electron mainly interacts with background electrons(CFL) through the Coulomb term,  $H_{int}=V_q\rho_q \rho^e_{-q}$ where $V_q=2\pi e^2/\epsilon q$, $\rho$ ($\rho_e$) the density of background (added)electron. Linked cluster approximation is appropriate here as well\footnote{Notice that the tunneling electron is described by a superposition of many bare CF states consisting of $2p$ particle-hole CF excitation and an inserted CF. As $p\rightarrow \infty$, it becomes such a complicated and high-energy object that it could hardly be scattered off or disassociated. Alternatively, the electron is a composite of CF and flux, and the flux will create a local trap potential on CFs in the CFL state, which ensures correct local density. As a result, a quasi-bound state is gapped from the extended state continuum. Thus, we expect the linked cluster approximation to be appropriate here. } and the electron's spectral function is now dressed by the density-density correlation function $\chi$\cite{PhysRevB.47.7312},

\begin{equation}
    \chi^r(q, u)\approx\frac{q}{2\pi}\frac{1}{i k_F \omega/q^2-\ell_B E_C}.
\end{equation}
which is expected to hold for $ u<{v_F^*} q$, below a cutoff scale $\Omega<\omega_c$ where ${v_F^*}$ is the Fermi velocity. As most of the contribution will come from small $q,u$ satisfy $u/E_C\sim (\ell_B q)^3\ll 1$, it's a good approximation to just integrate over $q$ from zero to infinity and $u$ from zero to the cutoff scale $\Omega$, we arrive at (details of the calculation below are given in Appendix  F), 
\begin{equation}
\begin{aligned}
        f_e(u)&=-\frac{1}{u^2} \int_0^\infty  \frac{q \mathrm{~d} q}{(2 \pi)^2} V_q V_{-q} \operatorname{Im}\left[\chi^r(q, u)\right]\\
        &=\frac{E_C^{\frac{1}{2}}}{4\sqrt{2}}u^{-\frac{3}{2}},\\
\end{aligned}
\end{equation}
and
\begin{equation}
    \begin{aligned}
        F_e(t)&=-\int_0^{\Omega} \mathrm{d} u f_e(u)\left(1-e^{-i u t}-iut\right).\\
    \end{aligned}
\end{equation}
Notice most of the contribution comes from low energy behavior; the limit, $\Omega/E_C\to\infty $, is well defined, and we get
\begin{equation}
    F_e(t)=-\sqrt{i\omega_0 t}+i\Bar{\omega}t.
\end{equation}
where $\omega_0=\pi E_C/8$, $\Bar{\omega}=\sqrt{E_C \Omega/8}$. 
The spectral function (of the particle side) is given by the Fourier transform of $\frac{1}{2}\exp(F_e(t))$ (a factor of one half is inserted as the total overlap on the particle side $\sum_\alpha O_\alpha^+=1/2$ as a result of particle-hole symmetry at filling $1/2$)
\begin{equation}
    A_+(\omega+E_e-\Bar{\omega})=\frac{1}{2}\Theta(\omega)\sqrt{\pi}e^{-\frac{\omega_0}{4\omega}}\left(\frac{\omega_0}{\omega}\right)^\frac{3}{2},
\label{elec_spec}
\end{equation}
where $E_e$ is the menafield energy of inserting an election into the CFL. 
%\ASc{Is $E_c$ the value of the energy within mean field?}
%(in the lowest Landau level Hilbert space). \ASc{It is not clear what you mean by that}. 
Eq.\ref{elec_spec} is valid from low energy up to the cutoff energy scale $\Omega$ where the form of density-density correlation function changes. {The hole part of the spectral function may be obtained in the same way. In both the electron and hole parts there are constants that we are not able to evaluate ($E_e$ and its corresponding $E_h$). If they are taken to be equal, the spectral function is particle-hole symmetric.} 
%, and the whole electron spectral function is going to be symmetric around zero energy \XYc{changed here} if we assume the ground state is particle hole symmetric\footnote{This particle hole symmetry is hidden or even absent in some low energy theory proposed for CFL\cite{PhysRevB.92.165125}}. 

\begin{figure}
    \centering
    \hspace{-0.75cm}
    \includegraphics[width = 250pt]{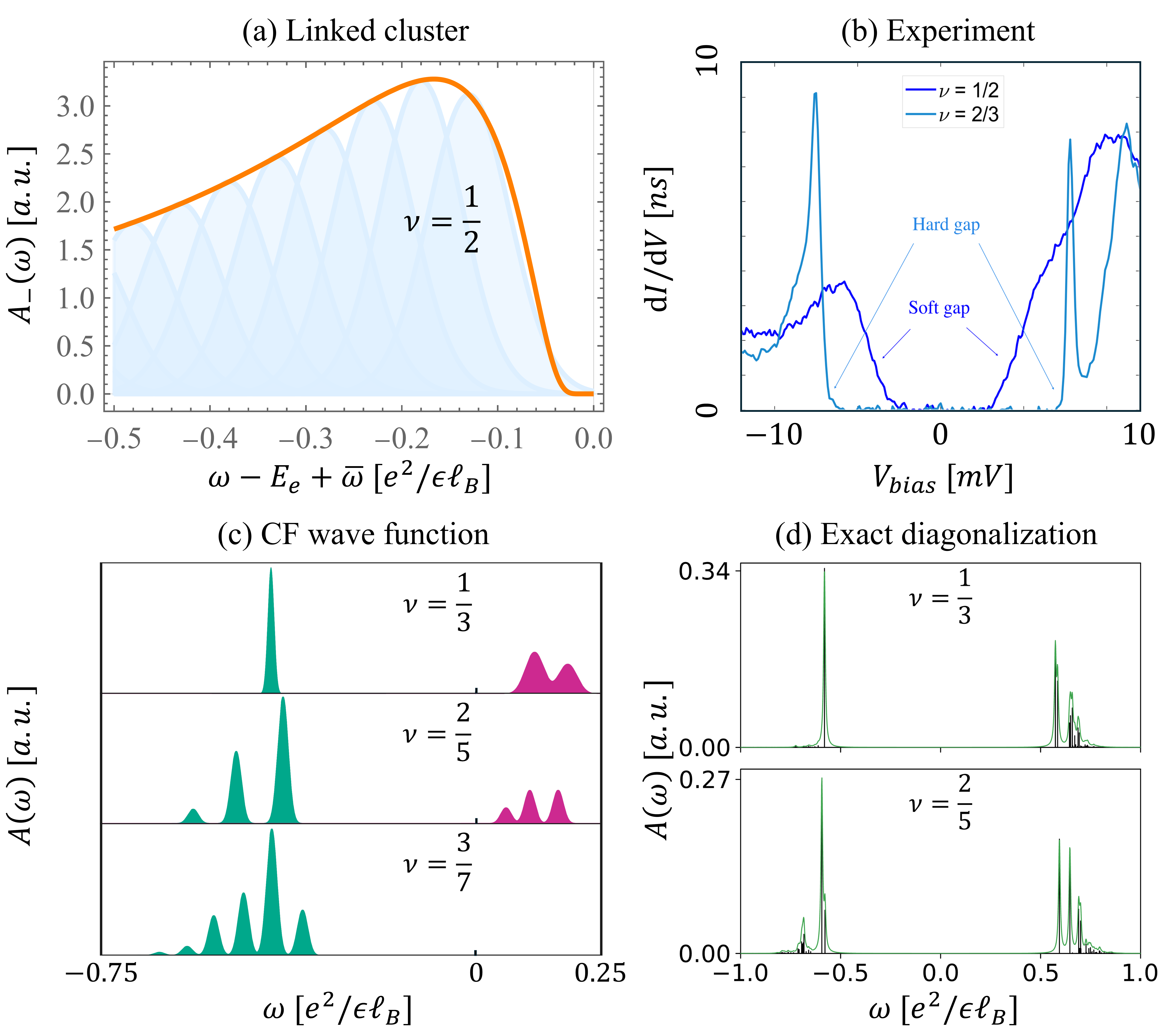}
    \caption{(a) Hole side of Electron spectral function (orange) at filling $\nu=\frac{p}{2p+1}$ for $p\gg 1$ based on linked cluster approximation. The most prominent feature is the sharp edge on the low energy side and the algebraic decay long tail on the high energy side. The shadow blue Gaussians illustrate different CF states' dressed spectral functions weighted by the corresponding overlap coefficients. Summation of these CF spectral functions should give us the half-filling result.
    (b) Scanning tunneling microscope measurement of FQH state at different filling factors in bilayer graphene system, figure reproduced from Hu et al.\cite{hu2023highresolution}.
    (c) Spectral function obtained from CF wave function based method, figure reproduced from Gattu et al. \cite{gattu2023stm}.
    (d)Electronic spectral function obtained from exact diagonalization calculation on a sphere (codes available at \cite{DVN/PJK5G8_2024}, modified based on codes developed by Ryan Mishmash\cite{halled2017}). We show results for $N=9$ electrons with monopole strength $Q=12$, which is the finite size realization of $\nu=1/3$, and $N=10$ with $Q=10.5$ which is the finite size realization of $\nu=2/5$. In the calculations, we use chord distance as the distance between electrons. A finite lifetime, $0.005E_C$, and a charging energy $\mu_\pm=0.5E_C$ have been introduced. More details about the exact diagonalization calculation can be found in Appendix G.
    }
    \label{half_filling}
\end{figure}

The hole side electron spectral function is shown in Figure.\ref{half_filling}(a), and the particle side is symmetric around zero frequency. The most prominent feature of the electron spectral function is the asymmetric behavior between low energy and high energy. We have a sharp cutting edge for low energy, whereas, for high energy, we have an algebraic decay tail. 
We remark that this asymmetry between low/high energy edge of the spectral function is observed in experiments\cite{farahi2023broken,hu2023highresolution}(Fig.\ref{half_filling}(b)), CF wave function based simulations\cite{gattu2023stm,pu2023fingerprints} (Fig.\ref{half_filling}(c)), and in exact diagonalization calculations (Fig.\ref{half_filling}(d)). We will discuss the experimental results in more detail later, and further discussions on these numerical methods can be found in Appendix G. The behavior at lowest energy was obtained before\cite{Halperin1993, PhysRevB.50.8078,levitov1997semiclassical}, whereas the power law decay behavior at the intermediate energy is proposed here for the first time to the best of our knowledge. 
Combining the electron spectral function we get here from the electron perspective and our previous results based on the CF perspective, we notice Eq.\ref{elec_spec} gives us information about the overlap $O_\lambda^\pm$ in the sense that it envelops the infinite Gaussian peaks we find in CF perspective. We expect Eq.\ref{elec_spec} to be a good approximation for $p$ not too small.

We now discuss the energy of the electron spectral function. As proved in Appendix  B, the first moment of the dressed spectral function stays unchanged at the bare value under linked cluster approximation. Eq.\ref{elec_spec} is normalizable, whereas the first moment of it diverges as $\sqrt{\Omega}$. This implies, not surprisingly, that Eq.\ref{elec_spec} is not UV complete, and it has to change to some other behavior at higher energy. In practice, we expect Eq.\ref{elec_spec} to capture the shape of the spectral function well below the cyclotron energy $\omega_c$.
Moreover, it is well known that $A_+(\omega)$ has to vanish for $\omega<0$ (as we are at zero temperature). This fact implies that we need $E_e-\Bar{\omega}\geq 0$ for our analysis to be self-consistent. While the pseudo-gapped spectral functions ($E_e-\Bar{\omega}= 0$) were obtained using similar approaches\cite{Halperin1993, PhysRevB.50.8078}, our analysis does not rule out the scenario of a true gap below the "pseudo-gapped" spectral function ($E_e-\Bar{\omega}> 0$), which is consistent with the expectation based on parton construction of the CFL theory. The latter scenario leads to the unusual situation that although the CFL state is compressible, adding a \textit{\textbf{local}} electron to a CFL state has zero overlap with a non-zero energy range above the Fermi energy. This does not happen for an electronic Fermi liquid, and reflects the CFL's strongly correlated nature. 
%In this sense, one might use orthogonal catastrophe to explain why electrons are gapped in CFL.

To summarize the above results, let's consider how the electron spectral function evolves if we decrease the filling factor from $1/2$. At filling factor $1/2$, we expect to see a symmetric spectral function described by Eq.\ref{elec_spec} on both sides. Now, imagine we decrease the filling factor gradually. At the filling factors where no quantization shows, we expect the system to be close to a CFL, and the corresponding spectral function is similar to the results at $1/2$. As the filling gets to FQHE states,   the width of the dressed CF peaks shrinks. Then, we expect the continuum incoherent spectral function Eq.\ref{elec_spec} to be split gradually and demonstrate directly the Landau quantization of CFL.

We close by discussing the relevant experimental results. Recently, high-quality scanning tunneling microscope(STM)  measurements were performed in fractional quantum Hall states on mono-layer graphene\cite{farahi2023broken} and bilayer-graphene\cite{hu2023highresolution}. These measurements,  Fig.\ref{half_filling}(b) for example, allow direct comparison with our analysis. Since STM is a probe that is local in space, the differential conductance data obtained by STM could be thought of as probing $A(QQQ,\omega)$ below the cyclotron frequency $\omega_c$. In these experiments, several clear incompressible states are observed at fractional fillings such as $1/3$, $2/3$ in single-layer graphene, and higher fillings such as $3/5$ in bilayer graphene. 
In bilayer graphene, sharp coherent peaks are observed in FQH states at small $p$ corresponding to filling factors such as $1/3$, $2/5$ $3/5$, and $2/3$ and are found to merge into a broad continuum for $p\geq 3$. As shown in Fig.\ref{half_filling}(b), the observed spectrum around $1/2$ is very similar to our result Eq.\ref{elec_spec}, which has a sharp edge on the low energy side and decays slowly on the high energy side. Deviating from half-filling, clear quantization behavior is observed in the experiment as the broad continuum starts to split into sharp peaks, which is consistent with our analysis. 

We may use the electronic excitation results to extract information about the underlying topological order. For example, we know from meanfield CF theory and numerical calculations (Fig.\ref{half_filling}(c)(d)) that there is only one peak in the hole side excitation spectrum in filling $\nu=1/3$, and more than one peak on the electron side. If the $\nu=2/3$ state is the particle-hole conjugate of the $\nu=1/3$ state, then the roles of the electron and hole sides should be exchanged. However, in the experiment on single-layer graphene\cite{farahi2023broken}, almost the same spectrum is observed at filling $-1+1/3$ and $-1+2/3$, and specifically, both of them have only one peak on the hole (negative energy) side (see Fig. \ref{half_filling}(b) for the $\nu=2/3$ spectrum). { Therefore, our analysis raises the possbility that these states might be described by the same $\nu=1/3$ topological order. And more precisely, the electronic excitation spectrum may suggest that this $-1+2/3$ state is not characterized by the P-H conjugate of $\nu=1/3$ but instead by two copies of $\nu=1/3$, a direct product of two $\nu=1/3$ Laughlin state formed by different flavors of  electron. }

On the other hand, we note several differences between our results and the experimental observations.
First, the peaks at higher energy do not correspond to smaller spectral weights. We suspect this is because of the additional electronic structure of the tip since, in the experiment, the particle and hole side of excitation do not correspond to the same spectral weight even for $\nu=1/2$. Second, our theory in the clean limit fails to predict that the discrete coherent peak will merge into a continuum for $p\geq 3$, in contrast to our expectation $p_c\approx 9$ based on large $p$ analysis, which implies the necessity of including additional incoherent scattering channels such as impurities and scattering process off the meanfield CF states (which is ignored in linked cluster approximation) to describe the experimental result accurately. 
%\XYc{If we take the vertex to be the normalized one, $\omega_c^*$, we expect $p_{merge}\sim\pi^5/2\sim150$ for the current cutoff we used, $1/R_\Delta$, of small $q$. However, this cutoff is not accurate to the level $O(1)$, say we could take it to be $O(1)/R_\Delta$ and $p_{merge}\sim \pi^5/2/O(1)^5$ }

In conclusion, our analysis merges the gap between the theoretical expectations of sharp coherent peaks and broad incoherent continuums in the electronic excitation spectrum of FQH states. Starting from meanfield approximation and considering the effect of fluctuating gauge field, we demonstrate that the two seemingly contradictory behaviors, sharp peaks and incoherent continuum, have the same physical origin and evolve from one to another even in the absence of impurities. We give an explicit expression for the electron spectral function at half-filling, which refines previous results\cite{Halperin1993} by extending the evaluated spectral function from low energy limit to intermediate energy scale and demonstrate the possibility of a clean gap below the pseudogap behavior.  {We find that our analysis is able to capture the dominant features in current experimental observations and numerical calculations.} Our analysis is useful for understanding the single electron spectroscopy experiments in FQH systems and guiding insights about underlying topological orders. It is possible to generalize our approach to describe the electronic excitations of other quantum-ordered states, for example, those which might be described by partons interacting with emergent gauge field\cite{wen2004quantum,sachdev2023quantum}.

\textit{Acknowledgement}. 
We thank Bo Han, Ali Yazdani, and Felix von Oppen for instructive discussions and the Free University in Berlin for their hospitality.
AS was supported by grants from the ERC under the European Union's Horizon 2020 research and innovation programme (Grant Agreements LEGOTOP No. 788715), the DFG (CRC/Transregio 183, EI 519/71), and the
ISF Quantum Science and Technology (2074/19).

%\newpage
\onecolumngrid
\setcounter{equation}{0}
\setcounter{figure}{0}
\renewcommand{\theequation}{A\arabic{equation}}
\renewcommand{\thefigure}{A\arabic{figure}}
\section{Appendix}
\subsection{A. Details about symmetry constraint on CF states}

In this section, we review the symmetry constraint used by Jain\cite{PhysRevLett.94.186808} to obtain the lowest energy CF state corresponding to adding/removing an electron from an FQH state in Jain's sequence, and we generalize this approach by using the angular momentum operators to obtain possible CF states at higher energy.
We consider a fractional quantum Hall state in Jain sequences at filling $\nu=p/(2\Tilde{\phi}p+1)<1/2$ with $p$ and $\tilde{\phi}$ positive integers, where $N$ electrons (assumed to be fully spin-polarized) live on a sphere with a monopole of strength $Q=\left ( \tilde{\phi}+\frac{1}{2p}\right ) N-\tilde{\phi}-\frac{p}{2}$ inside. This setup is the finite system version for filling factor $\nu=p/(2\Tilde{\phi}p+1)$ (we will limit ourselves to the $1>\nu>0$ range and zero-temperature limit for simplicity). 

In CF language, the CFs feel an effective magnetic monopole strength $Q^*=Q-\tilde{\phi}(N -1)=\frac{N}{2p}-\frac{p}{2}$ at meanfield level. The original electronic problem at fractional filling is therefore mapped to a composite fermion problem at integer filling, and the ground state $\ket{g}_N$ is described by $\ket{g}_N^{CF}$ in composite fermion language with $p$ filled CF Landau levels($\Lambda$-levels). Fig.\ref{CF_config}(a) shows an illustration of $\nu=1/3$ case.
Adding or removing an electron will change the effective monopole strength to $Q_\pm^*=Q-\tilde{\phi}(N_\pm -1)=Q^* \mp \tilde{\phi}$, as shown in Fig.\ref{CF_config}(b) and (c) respectively where the cross  sign emphasis the change of the degeneracy of the $\Lambda$-levels as a result of the added electron. An alternative way to understand this is considering the Streda formula, in which an electron could be considered as a CF attached with $2\tilde{\phi}$ flux in the opposite direction. We thus get $2\Tilde{\phi}p+1$ additional CF particles or holes on top of $p$ filled $\Lambda$-levels if we add or remove one electron. For simplicity, we will limit ourselves to $\tilde{\phi}=1$ cases from now on.

It is instructive to consider angular momentum operators here since we have full rotational symmetry\cite{PhysRevLett.71.424,Halperin1993, PhysRevLett.94.186808}.
The angular momentum operators could be written explicitly using
\begin{equation}
    \begin{aligned}
    \label{ang}
        L_z&=\sum_{j=Q}^\infty \sum_{m=-j}^j m c_{Qjm}^\dagger c_{Qjm}\\
    L_+&=\sum_{j=Q}^\infty \sum_{m=-j}^j e_{j,m}^+ c_{Qjm+1}^\dagger c_{Qjm}\\
    L_-&=\sum_{j=Q}^\infty \sum_{m=-j}^j  e_{j,m}^- c_{Qjm-1}^\dagger c_{Qjm}
    \end{aligned}
\end{equation}
with $e_{j,m}^\pm=\sqrt{j(j+1)-m(m\pm 1)}$ and $L_x=( L_+ +  L_- )/2$ and $L_y=( L_+ -  L_- )/2i$. $c_{Qjm}^\dagger$($c_{Qjm}$) is the electron creation (destruction) operator that creates (destroys) an electron in the $\left(j-Q\right)$th Landau level with ($\hbar=1$) total angular momentum $j$ and z-direction angular momentum(z-angular momentum) $m$.
This representation is reasonable because $L_z$ gives the total z-angular momentum, and these operators have the correct commutation relation. There's no additional constant, as one can verify using single-particle states.

In CF language, angular momentum operators are represented in the same form at meanfield level using CF operators $f_{Q^*jm}^\dagger$($f_{Q^*jm}$) that create(destroy) a CF in the $\left(j-Q^*\right)$th $\Lambda$-level with total angular momentum $j$ and z-direction angular momentum $m$. The meanfield CF representation of angular momentum operators is given by the similar form
\begin{equation}
    \begin{aligned}
        L_z&=\sum_{j=Q^*}^\infty \sum_{m=-j}^j m f_{Q^*jm}^\dagger f_{Q^*jm}\\
    L_+&=\sum_{j=Q^*}^\infty \sum_{m=-j}^j e_{j,m}^+ f_{Q^*jm+1}^\dagger f_{Q^*jm}\\
    L_-&=\sum_{j=Q^*}^\infty \sum_{m=-j}^j  e_{j,m}^- f_{Q^*jm-1}^\dagger f_{Q^*jm},
    \end{aligned}
\end{equation}
and could be justified similarly to the electron one.  We point out that although this CF representation is strictly valid only in the meanfield limit, it gives the correct result for states that do not even have a homogeneous density. For example, the largest angular momentum of the lowest energy particle-hole excitation is given correctly by pumping the CF with the smallest $L_z$ in the highest occupied $\Lambda$-level to the next $\Lambda$-level with the largest $L_z$, as one could see from the exact diagonalization calculation.
\begin{figure}
    \centering
    \includegraphics[width = 480pt]{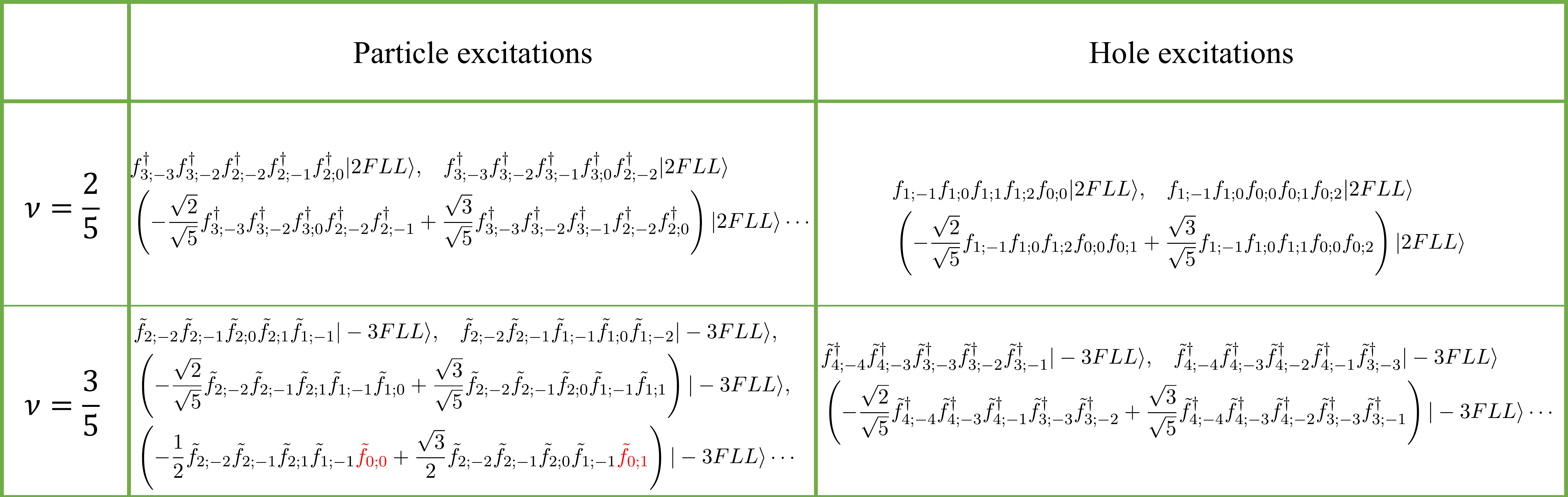}
    \caption{Low energy CF states corresponding to particle(hole) electronic excitations $c^\dagger_{QQQ}\ket{g}$
    ($c_{QQQ}\ket{g}$). $f_{i;j}^\dagger$($\tilde{f}_{i;j}^\dagger$) is the creation operator of CF in the $i$th $\Lambda$-levels with the $i+j-1$th largest(smallest) $z$-angular momentum. $|nFLL\rangle$ shows the ground state of  CFs at specific fillings with $n$ filled $\Lambda$-levels. The minus sign of $n$ emphasizes the CFs feel an effective magnetic field opposite to the external magnetic field. The red operators in some CF states indicate that these CF states involve CF excitations in the lowest $\Lambda$-level for $1/2<\nu<1$. The existence of such states \textbf{\textit{break}} the particle-hole symmetry between the particle-hole conjugate FQH states. (The coefficients in superpositioned CF states are obtained in large monopole strength limit.)}
    \label{cf_state_append}
\end{figure}
We assume the ground state $\ket{g}_N$ has zero angular momentum,
\begin{equation}
    \begin{aligned}
        \mathbf{L}^2\ket{g}_N&=0\ket{g}_N
    \end{aligned}
\end{equation}
where $\mathbf{L}^2=| \vec{\mathbf{L}}|^2$.
This condition is known to be true for Laughlin states, and it's easy to verify that $\ket{g}_N^{CF}$ indeed has zero angular momentum. As noted before, $c_{QQm}^\dagger\ket{g}_N$ are eigenstates of both $\mathbf{L}^2$ and $L_z$
\begin{equation}
    \begin{aligned}
    \label{sym_cond}
        \mathbf{L}^2\,c_{QQm}^\dagger\ket{g}_N&= Q(Q+1)\,c_{QQm}^\dagger\ket{g}_N\\
        L_z \;c_{QQm}^\dagger\ket{g}_N&= m\,c_{QQm}^\dagger\ket{g}_N.
    \end{aligned}
\end{equation}
This could be understood via addition rules of angular momentum(Clebsch-Gordan coefficients), $\ket{Q,m}\otimes\ket{0,0}=\ket{Qm}$, and is further justified by directly evaluating the commutator $[\textbf{L}^2,c_{QQm}^\dagger]$.

We start by considering the hole excitation $A_-(QQQ,\omega)$. In CF language, we look for CF states, $\ket{\beta_\lambda}_{N-1}^{CF}$, that have additional $2p+1$ CF holes on top of $p$ filled $\Lambda$-levels. Moreover, these states must have a component that has the same angular momentum quantum number as $c_{QQQ}\ket{g}_N$ to have a non-zero overlap. Jain uses constraint on $L_z$ and finds the allowed lowest energy state having $p+2$ and $p-1$ CF holes in the highest and the second highest occupied $\Lambda$-levels, respectively(see Fig.\ref{CF_config}(c) and Fig.\ref{link_cluster_diagram}(b) for example).

The particle excitation, $A_+(QQQ,\omega)$, could be predicted similarly. Now we have $2p+1$ CF on top of $p$ filled $\Lambda$-levels. However, since there's no bound on the largest $\Lambda$-level that the CFs could go(if we do not perform the lowest Landau level projection), we expect more peaks in the spectral function on the particle excitation side compared with the hole excitation side.

We generalize Jain's approach to get higher energy CF states with the correct total angular momentum. Notice the commutator between angular momentum operators and fermion creation(annihilation) operators
\begin{equation}
  \left[\textbf{L}^2,f_{Qjm}^\dagger\right]=j(j+1)f_{Qjm}^\dagger+2m f_{Qjm}^\dagger L_z + e_{jm}^+ f_{Qjm+1}^\dagger L_- - e_{jm}^- f_{Qjm-1}^\dagger L_+,
\end{equation}
one could find all the possible CF states satisfying the symmetry constraints. We summarise low energy ones for small $p$ states in Fig.\ref{CF_state} and Fig.\ref{cf_state_append}

\subsection{B. Linked Cluster approximation}
\begin{figure}
    \centering
    \includegraphics[width = 250pt]{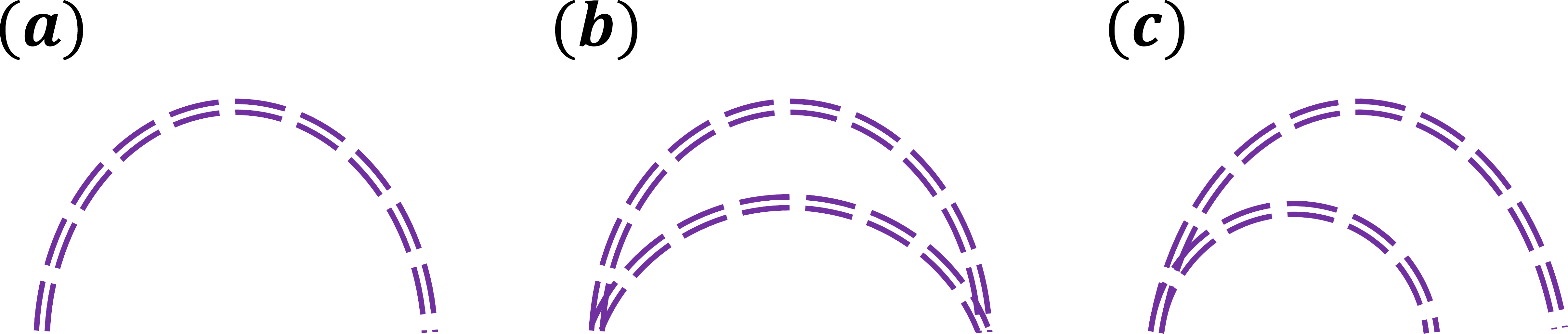}
    \caption{Linked clusters. (a) $F_2^1$, (b) $F_2^d$ and (c) $F_3$}
    \label{linked_cluster}
\end{figure}

In this section, we derive the expression for one electron's Green's function under linked cluster approximation. The C-S coupling reads
\begin{equation}
   \begin{aligned}
        \mathcal{V_{CS}}=\int_q \mathcal{V_{CS}}(q)&=\mathcal{V_{CS}}^0 + \mathcal{V_{CS}}^1 +\mathcal{V_{CS}}^d \\
        &=\int_q \left(-i a_{q}^0\int_k c^\dagger_{k+q}c_k- a_{q}^1\int_k \frac{k_y}{m^*} c^\dagger_{k+q}c_k+\int_{q'} a_{q-q'}^1 a_{q'}^1\int_k \frac{1}{2m^*} c^\dagger_{k+q}c_k\right)\\
   \end{aligned}
\end{equation}
where $c$s are the CF operators, we have used the notation that $q$ is in $\hat{x}$ direction, 0 and 1 for time and $y$ direction, respectively.
%Following the same spirit of RPA, the CS gauge field vertex with more than 2 CS field legs is assumed to be negligible \ASc{It does not exist in the bare level}. 
Therefore, we have only three kinds of linked clusters: two second-order linked clusters grouped into $F_2$ and one third-order linked cluster $F_3$,
\begin{equation}
    F_2(t)=F_2^1(t)+F_2^d(t)
\end{equation}
where $F_2^1(t)$ and $F_2^d(t)$, which  come from $\mathcal{V_{CS}}^1$ and $\mathcal{V_{CS}}^d$, are shown in Fig.\ref{linked_cluster} (a) and (b) respectively. 
We have,
\begin{equation}
   \begin{aligned}
        F_2^1(t)&=\frac{(-i)^2}{2}\int_q V_q^{\mu}V_{-q}^{\nu}\int_0^t dt_1\int_0^t dt_2 i D_{\mu\nu}(q,t_2-t_1)\\
        &=\frac{-i}{2}\int_q V_q^{\mu}V_{-q}^{\nu}\int_0^t dt_1\int_0^t dt_2 \left(\Theta(t_2-t_1)D_{\mu\nu}^> (q,t_2-t_1)+\Theta(t_1-t_2)D_{\mu\nu}^< (q,t_2-t_1)\right)\\
        &=-\frac{i}{2}\int_q V_q^{\mu}V_{-q}^{\nu}\int \frac{d\omega}{2\pi} \frac{1}{\omega^2}(1-e^{-i\omega t}) D_{\mu\nu}^> (q,\omega)+\int d\omega \frac{1}{\omega^2}(1-e^{i\omega t})D_{\mu\nu}^< (q,\omega)\\
        &=\int_q V_q^{\mu}V_{-q}^{\nu}\int \frac{d\omega}{2\pi} \frac{1}{\omega^2}\bigg\{(1-e^{-i\omega t}) \left(1+n_B(\omega)\right)+(1-e^{i\omega t})n_B(\omega)\bigg\}\Im[D_{\mu\nu}^r(q,\omega)]\\
        &=\int_q V_q^{\mu}V_{-q}^{\nu}\int \frac{d\omega}{2\pi} \frac{1}{\omega^2}(1-e^{-i\omega t}) \Im[D_{\mu\nu}^r(q,\omega)]\\
   \end{aligned}
   \label{derivation_lc}
\end{equation}
where $V_q$ is going to be the transition amplitude sandwiched by the CF fermion states with $N\pm1$ electrons. In the derivation, we drop linear in $t$ terms(constant energy shift) in the third line(we will come back to this energy shift at the end of this section), use the fluctuation-dissipation theorem in the fourth line, and take zero temperature limit finally (giving $n_B=1$). Similarly,
\begin{equation}
     F_2^d(t)=-2\int_{qq'}V^{d}_q V^{d*}_q \int\frac{d\omega d\omega'}{4\pi^2} \frac{1}{(\omega+\omega')^2}(1-e^{-i(\omega+\omega')t}) \Im[D^r_{11}(|q-q'|,\omega)]\Im[D^r_{11}(q',\omega')].
\end{equation}
Noteworthy, the integral over $q'$ is 2D. For $F_3$, as illustrated in Fig.\ref{linked_cluster}(c), we get
\begin{equation}
    \begin{aligned}
        F_3(t)&=\frac{(-i)^3}{3!}3\int_{qq'} V_{q-q'}^{1*}V_{q'}^{1*}V^{d}_q\int_0^t dt_1 \int_0^t dt_2 \int_0^t dt_3 D_{11}(q-q',t_2-t_1)D_{11}(q',t_3-t_1)\\
        &=-2\int_{qq'} V_{q-q'}^{1*}V_{q'}^{1*}V^{d}_q \int\frac{d\omega d\omega'}{4\pi^2\omega\omega'}\Im[D^r_{11}(q-q',\omega)]\Im[D^r_{11}(q',\omega')]\\
        &\qquad\qquad\times\bigg\{\frac{1}{\omega+\omega'}(1-e^{-i(\omega+\omega')t})-\frac{1}{\omega}(1-e^{-i\omega t})
        -\frac{1}{\omega'}(1-e^{-i\omega' t})\bigg\}
    \end{aligned}
\end{equation}
using a similar method. In practice, $F_2^1(t)$ contributes almost all the effect since energy-momentum conservation is hard to satisfy for $F_2^d$ and $F_3$.

\begin{figure}
    \centering
    \includegraphics[width=0.45\linewidth]{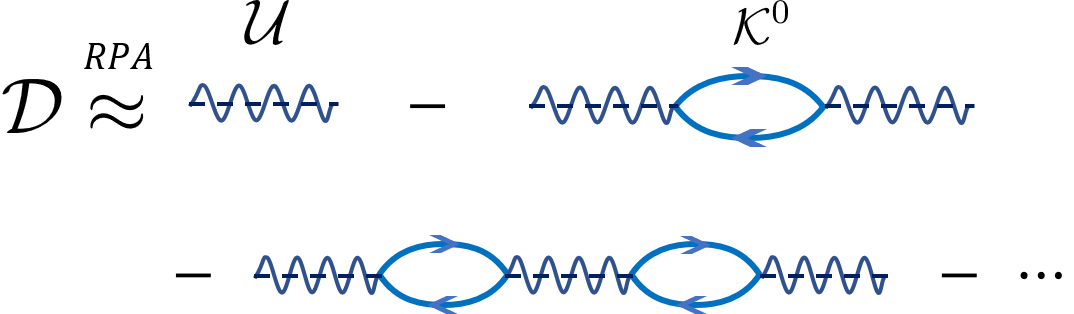}
    \caption{illustration of the C-S propagator}
    \label{dcs}
\end{figure}

Now we come back to the discussion of the constant energy shift we omit in Eq.\ref{derivation_lc}. Below, we prove that the first moment of the dressed spectral function remains unchanged; in other words, the averaged energy is unchanged by the linked cluster approximation. The full expression at zero temperature reads,
\begin{equation}
   \begin{aligned}
        F_2^1(t)&=\int_q V_q^{\mu}V_{-q}^{\nu}\int \frac{d\omega}{2\pi} \left[\frac{1}{\omega^2}(1-e^{-i\omega t})-\frac{it}{\omega}\right] \Im[D_{\mu\nu}^r(q,\omega)]\\
        &=\int \frac{d\omega}{2\pi} \left[-f(\omega)(1-e^{-i\omega t})+it\omega f(\omega)\right]\\
   \end{aligned}
   \label{f2t}
\end{equation}
we now evaluate the first moment of the spectral function \textit{without} including the constant energy shift, using Eq.\ref{derivation_lc},
\begin{equation}
    \begin{aligned}
        \langle E \rangle&=\int \frac{\dd \omega}{2\pi}\omega A(\omega)\\
        &=\int \frac{\dd \omega}{2\pi}2\pi e^{-\int \frac{\dd u}{2\pi} f(u)}\left[\omega\delta(\omega)+\sum_{l=1}^\infty \omega \frac{\delta (\omega-\sum_{i=1}^l u_i)}{l!}\prod_{i=1}^l\int\frac{\dd u_i}{2\pi}f(u_i)\right]\\
        &=\int\frac{\dd u}{2\pi}uf(u)
    \end{aligned}
\end{equation}
where we used Eq.\ref{expression_spec} and the fact that the unperturbed state is an eigenstate(we assume the corresponding eigenenergy to be zero). One immediately realizes that the change in averaged energy caused by linked cluster Eq.\ref{derivation_lc} compensates with the energy shift term we omitted. Therefore, the linked cluster approximation keeps the first moment of the spectral function unchanged. In other words, the "center" of the spectral function stays unchanged before and after taking into account linked cluster processes.

\subsection{C. RPA density response and dressed CF meanfield states for small $p$}
In this section, we review the calculation of the dressed CS propagator following Simon and Halperin's modified RPA method\cite{PhysRevB.48.17368}, and show the dressed spectral function(under linked cluster) of a single CF state for small $p$. Simon and Halperin's method provides a density response function $K$ satisfying Kohn's theorem. Consider a FQH state at $\nu=p/2p+1$ (we set $\ell_B=\hbar=c=e=1$), the conductivity matrix is given by
\begin{equation}
    s_n=\rho_n^{-1}=\frac{p }{2 \pi }\left[\begin{array}{cc}
i\left(\frac{\omega}{\Delta \omega_c^*}\right) \Sigma_0 & \Sigma_1 \\
-\Sigma_1 & i\left(\frac{\Delta \omega_c^*}{\omega}\right)\left(\Sigma_2+1\right)
\end{array}\right]
\end{equation}
where $\Delta \omega_c^*=\frac{\omega_c^*}{2p+1}$
\begin{equation}
    \Sigma_j=\frac{e^{-Y}}{p} \sum_{l=0}^{p-1} \sum_{m=p}^{\infty}\left\{\frac{l !}{m !} \frac{(m-l) Y^{m-l-1}\left[L_l^{m-l}(Y)\right]^{2-j}}{\left(\frac{\omega+i 0^{+}}{\Delta \omega_c^*}\right)^2-(m-l)^2}\left[(m-l-Y) L_l^{m-l}(Y)+2 Y \frac{d L_l^{m-l}(Y)}{d Y}\right]^j\right\}
\end{equation}
with $L_n^{\alpha}$ the associate Laguerre polynomial, $Y=\frac{2p+1}{2}q^2$. For calculation, we use the summed form for these $\Sigma_j$ as discussed in Simon's original paper. The density response $K$ and the propagator $D$ are obtained by the relations,

\begin{equation}
    \begin{aligned}
        D&=U-UKU\\
        K^{-1}&=T\rho T+V\\
        \rho&=\rho_n+\rho_{FL}+\rho_{CS}\\
        \rho_{FL}&=-\frac{i \omega\left(m_b/m^*-1\right)}{\Delta \omega_c^*}\left[\begin{array}{ll}
1 & 0 \\
0 & 1
\end{array}\right]\\
        \rho_{\mathrm{CS}} &\equiv T^{-1} C^{-1} T^{-1}=2 \pi  \tilde{\phi}\left[\begin{array}{cc}
0 & 1 \\
-1 & 0
\end{array}\right]\\
    \end{aligned}
\end{equation}
where
\begin{equation}
    T=e\left[\begin{array}{cc}
\frac{i \sqrt{i \omega}}{q} & 0 \\
0 & \frac{1}{\sqrt{i \omega}}
\end{array}\right]
\qquad
C=\left[\begin{array}{cc}
0 & \frac{i q}{2 \pi  \tilde{\phi}} \\
-\frac{i q}{2 \pi  \tilde{\phi}} & 0
\end{array}\right]
\qquad
V=\left[\begin{array}{cc}
v(q) & 0 \\
0 & 0
\end{array}\right]
\end{equation}
and
\begin{equation}
    U=V+C^{-1}
\end{equation}

We take $\omega_c^*=0.3\frac{e^2}{\epsilon \ell_B}=0.3E_C$ estimate from exact diagonalization results and use $m_b/m^*=1/\infty$ since we are in the lowest Landau level limit.

\begin{figure}
    \centering
    \hspace{-0.75cm}
    \includegraphics[width = 180pt]{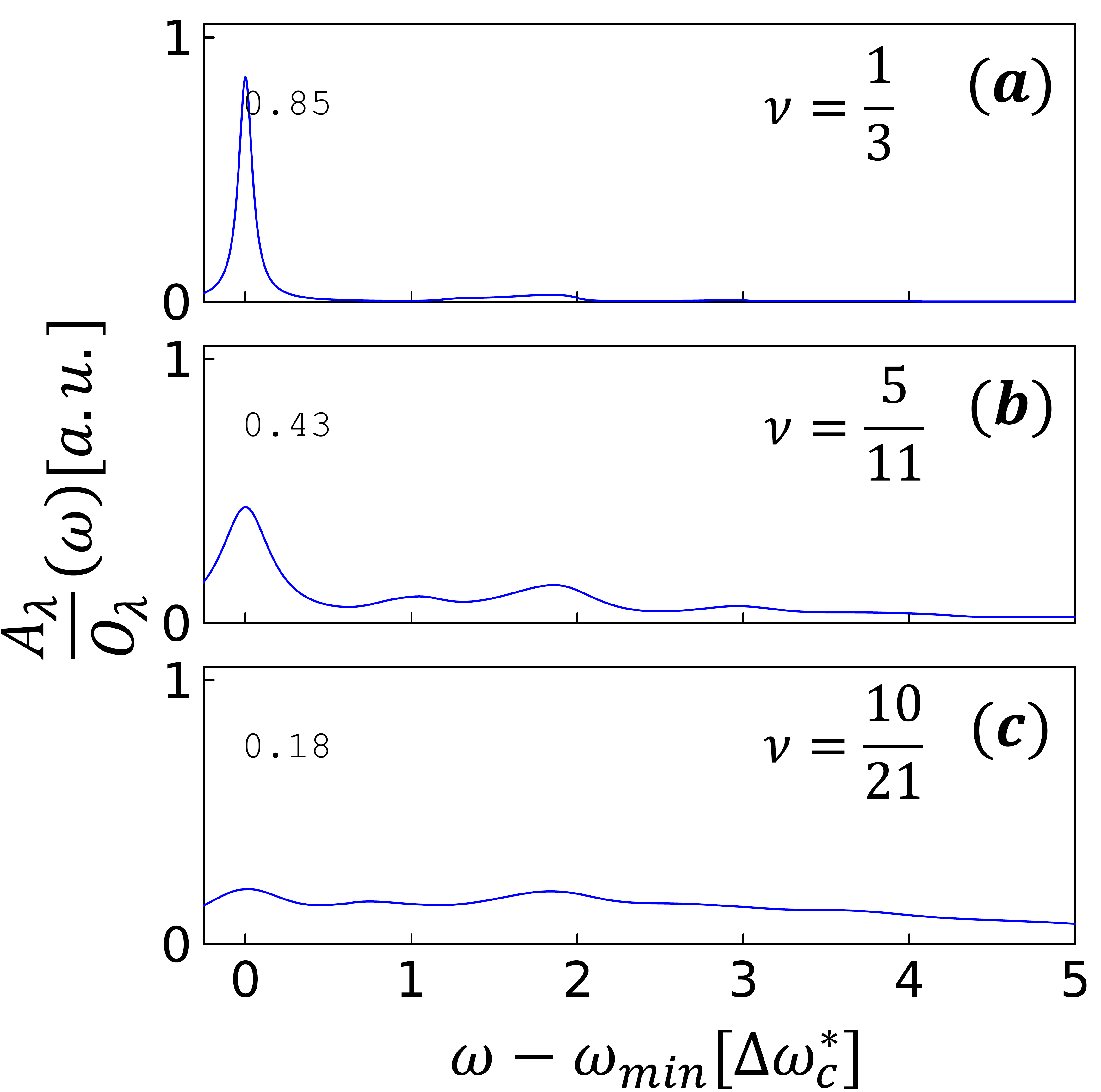}
    \caption{Dressed spectral function of a single meanfield CF state obtained under linked cluster approximation for different filling factors. The number around the peak shows the spectral weight of the lowest energy peak. A finite lifetime, $0.017\omega_c^*=(2p+1)\times0.017\Delta\omega_c^*$, is introduced for illustration. It's clear from the figure that for large $p$, the coherent peak evolves into a broad continuum as a result of gauge field fluctuation. The only physically relevant parameter we put in by hand in this calculation is the ratio between the effective mass and the Coulomb energy. We estimate this from the exact diagonalization calculation. The vanishing of sharp coherent peaks is an inevitable trend for FQH states close to compressible CFL states due to diminishing energy gaps.}
    \label{link_cluster_spectrum}
\end{figure}

We implement the linked cluster approximation by evaluating the dressed spectral function using $D_{\mu\nu}^r(q,u)$ obtained from the above modified random phase approximation\cite{PhysRevB.48.17368}. We evaluate the dressed spectral function of a \textit{single} peak predicted by meanfield CF theory in Fig.\ref{link_cluster_spectrum} for filling factors from $1/3$ to $10/21$.
For small $p$, the dressed spectral function of a single bare CF state is still dominated by an infinite sharp peak at the lowest energy. We, therefore, expect the spectral function to contain a series of sharp peaks for these small $p$ states. The exact filling factor at which the coherent peaks would disappear will depend on impurities, which is absent in our analysis.
%The spectral function is normalized such that the height of the lowest energy peak equals its spectral weight \ASc{you mean its spectral weight within mean field theory?}. 
As the filling factor approaches $1/2$, the gap of the gauge field fluctuation becomes smaller, and therefore, the bare CF state mixes more with the fluctuations and overlaps less with the original bare CF state. As a result, most of the spectral weight is distributed to higher energy (relative to $\Delta \omega_c^*$), and the lifetime becomes finite.

\subsection{D. Coupling Vertex}
We now evaluate the matrix element of this coupling to the CS gauge field using the LL basis in the rotational invariant gauge,
$$
\boldsymbol{A_S}=\frac{\boldsymbol{B} \times \boldsymbol{r}}{2}=\frac{B}{2}(-y, x, 0).
$$
Denote the charge of an electron to be $-e$. Here we set $eB/c=1$, $\hbar=1$, $m=1$ and specifically the magnetic length $\ell=1$. When evaluating the matrix element between CF states, one should re-insert the unit. The Hamiltonian for 2D non-relativistic electrons under a perpendicular magnetic field could be written as 
$$
    H=\frac{1}{2}\left[\left(-\mathrm{i} \frac{\partial}{\partial x}-\frac{y}{2}\right)^2+\left(-\mathrm{i} \frac{\partial}{\partial y}+\frac{x}{2}\right)^2\right]
$$
Using 
$$
z=x-\mathrm{i} y=r \mathrm{e}^{-\mathrm{i} \theta}, \quad \bar{z}=x+\mathrm{i} y=r \mathrm{e}^{\mathrm{i} \theta}
$$
we get 
$$
H=\frac{1}{2}\left[-4 \frac{\partial^2}{\partial z \partial \bar{z}}+\frac{1}{4} z \bar{z}-z \frac{\partial}{\partial z}+\bar{z} \frac{\partial}{\partial \bar{z}}\right] .
$$
We define the following sets of ladder operators:
$$
\begin{aligned}
b & =\frac{1}{\sqrt{2}}\left(\frac{\bar{z}}{2}+2 \partial\right), \qquad
b^{\dagger} & =\frac{1}{\sqrt{2}}\left(\frac{z}{2}-2 \bar{\partial}\right), \\
a^{\dagger} & =\frac{1}{\sqrt{2}}\left(\frac{\bar{z}}{2}-2 \partial\right), \qquad
a & =\frac{1}{\sqrt{2}}\left(\frac{z}{2}+2 \bar{\partial}\right),
\end{aligned}
$$
with 
$$
\partial=\frac{1}{2}\left( \partial_x+i\partial_y \right)
\qquad
\bar{\partial}=\frac{1}{2}\left( \partial_x-i\partial_y \right)
$$
The Hamiltonian is in the form 
\begin{equation}
\nonumber
    H=a^{\dagger} a+\frac{1}{2},
\end{equation}
and commute with the angular momentum operator,
\begin{equation}
\nonumber
    \begin{aligned}
L_z & =-\mathrm{i} \hbar \frac{\partial}{\partial \theta} \\
& =-\hbar\left(z \frac{\partial}{\partial z}-\bar{z} \frac{\partial}{\partial \bar{z}}\right) \\
& =-\hbar\left(b^{\dagger} b-a^{\dagger} a\right),
\end{aligned}
\end{equation}
The eigenstates, e.g. $\ket{N,m}$, are characterized by $H$ and $L_z$ with quantum number $N,m$ correspondingly. The fermionic part of $ \mathcal{V_{CS}}$ is some momentum shift operator in single-particle Hilbert space and could be written as 
\begin{equation}
    S_q^0= e^{i\hat{r}\cdot\bf{q}}
\end{equation}
and
\begin{equation}
    S_q^1= -e^{i\hat{r}\cdot\bf{q}}\hat{k_y}=e^{i\hat{r}\cdot\bf{q}}i\partial_y.
\end{equation}
Since
\begin{equation}
\begin{aligned}
    x&=\frac{1}{2}(z+\bar{z})&&=\frac{1}{\sqrt{2}}(a+a^\dagger+b+b^\dagger)\\
    y&=\frac{i}{2}(z-\bar{z})&&=\frac{i}{\sqrt{2}}(a-a^\dagger-b+b^\dagger)\\
    i\partial_y&=\;\;\;\;\partial-\bar{\partial}&&=\frac{1}{2\sqrt{2}}(b+b^\dagger-a-a^\dagger)
\end{aligned}
\end{equation}
%\ASc{Do you implicitly set $\ell_\Delta=1$? I think that you are not consistent between Eq. S.34 and S.40}
we get
\begin{equation}
    e^{ixq_x+iyq_y}=e^{\frac{i}{\sqrt{2}}[(q_x+iq_y)a+(q_x-iq_y)a^\dagger]}e^{\frac{i}{\sqrt{2}}[(q_x-iq_y)b+(q_x+iq_y)b^\dagger]}.
\end{equation}
We could simplify the above expression by redefining $\frac{q_x+iq_y}{q}a\rightarrow a$, etc. . The price we pay is a phase factor to the expression (but for this case, it is zero since $q$ is in $\hat{x}$ direction).
For single-particle states, the matrix element is easy to evaluate
\begin{equation}
    \bra{M,m}S_q^0\ket{N,n}=\bra{M,m}e^{\frac{i}{\sqrt{2}}q(a+a^\dagger)}e^{\frac{i}{\sqrt{2}}q(b+b^\dagger)}\ket{N,n}
\end{equation}
\begin{equation}
    \bra{M,m}S_q^1\ket{N,n}=\frac{1}{2\sqrt{2}}\bra{M,m}e^{\frac{i}{\sqrt{2}}q(a+a^\dagger)}e^{\frac{i}{\sqrt{2}}q(b+b^\dagger)}(b+b^\dagger-a-a^\dagger)\ket{N,n}
\end{equation}
using BCH formula
\begin{equation}
    e^{\frac{i}{\sqrt{2}}q(a+a^\dagger)}=e^{-\frac{q^2}{4}}e^{\frac{i}{\sqrt{2}}q 
a^\dagger}e^{\frac{i}{\sqrt{2}}q a}
\end{equation}
and a similar expression for $b$, we get, for $m\geq n$
\begin{equation}
\begin{aligned}
    t_{mn}&=e^{-\frac{q^2}{4}}\bra{0,m}e^{\frac{i}{\sqrt{2}}q 
b^\dagger}e^{\frac{i}{\sqrt{2}}q b}\ket{0,n}\\
&=e^{-\frac{q^2}{4}}\left(\frac{iq}{\sqrt{2}}\right)^{m-n}\left(\frac{n!}{m!}\right)^{\frac{1}{2}}L_n^{(m-n)}(\frac{q^2}{2}),
\end{aligned}
\end{equation}
for $m<n$,
\begin{equation}
\begin{aligned}
    t_{mn}&=e^{-\frac{q^2}{4}}\bra{0,m}e^{\frac{i}{\sqrt{2}}q 
b^\dagger}e^{\frac{i}{\sqrt{2}}q b}\ket{0,n}\\
&=e^{-\frac{q^2}{4}}\left(\frac{iq}{\sqrt{2}}\right)^{n-m}\left(\frac{m!}{n!}\right)^{\frac{1}{2}}L_m^{(n-m)}(\frac{q^2}{2}),
\end{aligned}
\end{equation}
where $L_n^{(\alpha)}(x)$ is the associated Laguerre polynomial. Now, we could express the matrix elements using $t_{mn}$(we put back the length unit $\ell_B$)
\begin{equation}
     \begin{aligned}
        \bra{M,m}\mathcal{V_{CS}}^0 (q)\ket{N,n}&=a_q^0 \;V^0_{q}(Mm;Nn)=-i a_q^0 \;t_{MN}t_{mn}\\
        \bra{M,m}\mathcal{V_{CS}}^1(q)\ket{N,n}&=a_q^1 \;V^1_{q}(Mm;Nn)\\
        &=-\frac{a_{q}^1}{m^*\ell_B}\frac{1}{2\sqrt{2}}\left(\sqrt{n}t_{MN}t_{mn-1}+\sqrt{n+1}t_{MN}t_{mn+1}-\sqrt{N}t_{MN-1}t_{mn}-\sqrt{N+1}t_{MN+1}t_{mn}\right)\\
        \bra{M,m}\mathcal{V_{CS}}^d(q)\ket{N,n}&=\int_{q'} a_{q-q'}^1 a_{q'}^1 \;V^{d}_{q}(Mm;Nn) =\int_{q'} a_{q-q'}^1 a_{q'}^1 \frac{1}{2m^*}\;t_{MN}t_{mn}\\.
     \end{aligned}
     \label{coupling_matrix}
\end{equation}
Noteworthy, when we evaluate the CF vertex, we need to use the effective magnetic field for CFs, $B\rightarrow\Delta B$. 

The coupling vertex $V_q^0$($V_q^1$) is the sum of $V_{CS}$ sandwiched by the meanfield CF states corresponding to adding or removing an electron. For the lowest energy hole excitation $\ket{\beta_{p}}$ in filling $\nu=p/(2p+1)$, we find the vertex to be 
\begin{equation}
     \begin{aligned}
        \bra{\beta_{p}}\mathcal{V_{CS}}^0 (q)\ket{\beta_{p}}&=\sum_{i=0}^{p+1}\bra{p-1,i}\mathcal{V_{CS}}^0 (q)\ket{p-1,i}+\sum_{i=0}^{p-2}\bra{p-2,i}\mathcal{V_{CS}}^0 (q)\ket{p-2,i}\\
        &=e^{-\frac{q^2}{2}}\left(L_{p-1}^{(0)}(\frac{q^2}{2})L_{p+1}^{(1)}(\frac{q^2}{2})+L_{p-2}^{(0)}(\frac{q^2}{2})L_{p-2}^{(1)}(\frac{q^2}{2})\right)
     \end{aligned}
     \label{summered_vertex}
\end{equation}
for large $p$ it becomes,
\begin{equation}
    \begin{aligned}
        \bra{\beta_{p}}\mathcal{V_{CS}}^0 (q)\ket{\beta_{p}}&\approx  -2i a_q^0 e^{-\frac{q^2}{2}}L_{p}^{(0)}(\frac{q^2}{2})L_{p}^{(1)}(\frac{q^2}{2})\\
        &\approx -i a_q^0\frac{2\sqrt{2p}}{q}J_0(\sqrt{p}q)J_1(\sqrt{p}q)
    \end{aligned}
\end{equation}
where $J_\alpha$ is the first kind of Bessel function, and we have used 
\begin{equation}
    L_n^{(\alpha)}(x)\approx n^\alpha e^{x / 2} \frac{J_\alpha(2 \sqrt{nx})}{\sqrt{nx}^\alpha}
\end{equation}
Similarly, we have 
\begin{equation}
        \bra{\beta_{p}}\mathcal{V_{CS}}^1 (q)\ket{\beta_{p}}
        \approx-\frac{2i p a_q^1}{m^* q\ell_B}\left\{J_0(\sqrt{p}q)J_2(\sqrt{p}q)-J_1(\sqrt{p}q)J_1(\sqrt{p}q)\right\}
        \label{vs1}
\end{equation}
We note that, for large $p$, the approximation above is a good approximation for both the electron and hole lowest energy CF states.

\subsection{E. Large $p$ limit}

In this section, we perform the calculation for large $p$ in which we could get an analytic expression for the shape of the spectral function up to the leading order in $p$. Since we've found the dominant contribution comes from $D_{11}$ for large $p$, we will focus on the contribution from this element. For large $p$, we expect the expression of $D_{11}$ at half filling is a good approximation as we will see later that the contribution mainly comes from wave vector $q R_\Delta \gg 1$ where $R_\Delta=\sqrt{2p(2p+1)}\ell_B$ is the cyclotron radius.

For $\nu=1/2$, the expression of $D_{11}$ was obtained\cite{PhysRevB.47.7312} under RPA approximation and found to be, for $\omega<{v_F^*} q$,
\begin{equation}
    D_{11}(q, \omega) \approx\left[i\frac{k_F}{2\pi} \frac{\omega}{q}-q \ell_B \frac{E_C}{8 \pi }\right]^{-1}
\end{equation}
where $k_F$ is the Fermi wave vector, and we have dropped the term with higher order in $q$ as we will find the contribution dominated by $q \ell_B\ll 1$. We thus obtain 
\begin{equation}
    \Im{D_{11}^r (q,\omega)}=-2\pi k_F \frac{q\omega}{\varphi^2 q^4+k_F^2 \omega^2}
\end{equation}
with $\varphi=\frac{\ell_B E_C}{4}$. 

Now we re-insert the length unit $\ell_\Delta$ for the coupling vertex between CF states and CS gauge field. The coupling matrix element $V_q^1$ for the lowest energy electron/hole excitation is approximated by (up to leading order in $p$)

\begin{equation}
    V_q^1 \approx\left\{\begin{aligned}
&i\frac{\Delta{\omega_c^*} \ell_\Delta p^2}{4} q \ell_\Delta = i\frac{ {v_F^*} R_\Delta q}{8}&&\text { for } \qquad \qquad q R_\Delta < c_v\approx 2 \\
&i\frac{2\sqrt{2}\Delta{\omega_c^*} R_\Delta}{\pi q^2\ell_\Delta^2} \approx i\frac{4\sqrt{2}  {v_F^*} p}{\pi R_\Delta^2 q^2} &&\text { for } \qquad c_v < q R_\Delta \ll R_\Delta / l_{\Delta} \\
&\text { negligible}&&\text{ for } \qquad  1\ll q l_{\Delta}
\end{aligned}\right.
\end{equation}
where we have used Eq.\ref{vs1} and the approximated expression for the Bessel function at large $p$, we get the cutoff value, $c_v\approx 2$, by directly plotting the asymptotic expression and the exact expression and comparing them. $\Delta{\omega_c^*}=\frac{\omega_c^*}{2p+1}$ is the meanfield effective CF cycrotron frequency and $\ell_\Delta=\sqrt{2p+1}\ell_B$.

One immediately realizes that the leading contribution will come from the intermediate wave vector region. We thus get the expression for the suppression factor $S(p)=\exp(-\int_0^{\infty}d u f(u))$ at large $p$ (at leading order in $p$)
\begin{equation}
    \begin{aligned}
        S(p)=&\exp \left(\int_{\tilde{\Delta}}^{\infty} \mathrm{d} u \frac{1}{u^2} \int \frac{q \mathrm{~d} q}{(2 \pi)^2} V_q^1 V_{-q}^1 \operatorname{Im}\left[D_{11}^r(q, u)\right]\right)\\
        \approx& \exp \left(-\int_{\tilde{\Delta}}^{\infty} \mathrm{d} u \frac{1}{u^2} \int_{\frac{c_v}{R_{\Delta}}}^{\frac{1}{\ell_\Delta}} \frac{q \mathrm{~d} q}{(2 \pi)^2} \abs{\frac{2\sqrt{2}i\Delta{\omega_c^*} R_\Delta}{\pi q^2\ell_\Delta^2}}^2 \frac{2\pi k_F q u}{\varphi^2 q^4+k_F^2 u^2}\right)\\
        =&\exp\Bigg(-\sqrt{\varphi}\abs{\frac{2\sqrt{2}\Delta{\omega_c^*} R_\Delta}{\pi \ell_\Delta^2}}^2\frac{ k_F}{40\pi q^5 u^{5 / 2}}\Bigg\{8 q^4 \sqrt{u}-2 \sqrt{2} q^5 \operatorname{\arctan}\left[1-\frac{\sqrt{2} q}{\sqrt{u}}\right]+2 \sqrt{2} q^5 \operatorname{\arctan}\left[1+\frac{\sqrt{2} q}{\sqrt{u}}\right]\\
        &+\sqrt{2} q^5 \ln \left[\frac{q^2-\sqrt{2} q \sqrt{u}+u}{q^2+\sqrt{2} q \sqrt{u}+u}\right]+2 u^{5 / 2} \ln \left[\frac{q^2-\sqrt{2} q \sqrt{u}+u}{u}\right]\\
        &+2 u^{5 / 2} \ln \left[\frac{q^2+\sqrt{2} q \sqrt{u}+u}{u}\right]\Bigg\}\Bigg|_{q=\frac{\sqrt{\varphi}}{R_{\Delta}}} ^{\frac{c_v\sqrt{\varphi}}{\ell_{\Delta}}}\Bigg|_{u=k_F \tilde{\Delta}} ^{\infty}\Bigg)\\
        \approx&\exp\Bigg(-\frac{2 {v_F^*}^2R_\Delta}{c_v \pi^3(2p+1)^2\ell_B^4 k_F \tilde{\Delta}^2}\Bigg)\\
        =&\exp\Bigg(-\frac{8(2p+1)\ln^2{(2p+1)}{\omega_c^*}^2}{c_v \pi^5 E_c^2}\Bigg)
    \end{aligned}
\end{equation}

Along the derivation, we used the relation $k_F\approx 1/\ell_B$, $k_F R_\Delta \approx 2p+1$ and introduced an energy cut off (since for any finite $p$ the state is gapped) which was evaluated\cite{PhysRevB.52.5890} under the same scheme to be $\tilde{\Delta}(p)\approx \frac{E_C\pi}{ 2(2 p+1)} \frac{1}{\ln (2 p+1)}$. We change the variable in the third line and keep the term up to leading order in $p$ in the fourth line. The suppression factor is dominated by low energy and long-distance behavior.

The above equation shows how the suppression evolves when $p$ increases. What's more, as we expected, there are no sharp coherent peaks in the electronic excitation spectrum for filling close to $1/2$.

%{\frac{1}{\ell_\Delta}} \dd q \frac{k_F R_\Delta}{2\pi^2}  \frac{q u}{\varphi q^4+k_F^2}

It is interesting to evaluate the distance between the lowest energy peak and the first moment of the spectral function. We have
\begin{equation}
    \begin{aligned}
        \langle E \rangle=&-\int_{\tilde{\Delta}}^{\infty} \mathrm{d} u \frac{1}{u} \int \frac{q \mathrm{~d} q}{(2 \pi)^2} V_q^1 V_{-q}^1 \operatorname{Im}\left[D_{11}^r(q, u)\right]\\
        \approx& \int_{\tilde{\Delta}}^{\infty} \mathrm{d} u \frac{1}{u} \int_{\frac{c_v}{R_{\Delta}}}^{\frac{1}{\ell_\Delta}} \frac{q \mathrm{~d} q}{(2 \pi)^2} \abs{\frac{2\sqrt{2}i\Delta{\omega_c^*} R_\Delta}{\pi q^2\ell_\Delta^2}}^2 \frac{2\pi k_F q u}{\varphi^2 q^4+k_F^2 u^2}\\        =&\sqrt{\varphi}\abs{\frac{2\sqrt{2}\Delta{\omega_c^*} R_\Delta}{\pi \ell_\Delta^2}}^2\frac{ 1}{24\pi q^3 u^{3 / 2}} \Biggl\{8 q^2 \sqrt{u} - 
  2 \sqrt{2} q^3 \arctan[1 - (\sqrt{2} q)/\sqrt{u}] + 
  2 \sqrt{2} q^3 \arctan[1 + (\sqrt{2} q)/\sqrt{u}]\\
  &+  4 u^{3/2} \arctan[1 - (\sqrt{2} \sqrt{u})/q] +   4 u^{3/2} \arctan[1 + (\sqrt{2} \sqrt{u})/q] \\&+   \sqrt{2} q^3 \ln[q^2 - \sqrt{2} q \sqrt{u} + u] -  \sqrt{2} q^3 \ln[q^2 + \sqrt{2} q \sqrt{u} + u])\biggr\}\Bigg|_{q=\frac{\sqrt{\varphi}}{R_{\Delta}}} ^{\frac{c_v\sqrt{\varphi}}{\ell_{\Delta}}}\Bigg|_{u=k_F \tilde{\Delta}} ^{\infty}\\
  \approx&\frac{8{\omega_c^*}^2}{c_v\pi^2 E_c}\ln [2p+1]
    \end{aligned}
\end{equation}

To get the overall shape of the spectral function in large $p$ limit, we use the complete expression for $F_2(t)$, Eq.\ref{f2t}, and consider the spectral function in the time domain,
\begin{equation}
\begin{aligned}
        A_\alpha(t)&=\exp(F_2(t)-iE_\alpha t)\\
        &=\exp \left(\int_{\tilde{\Delta}}^{\infty} \mathrm{d} u \frac{1}{u^2} \int \frac{q \mathrm{~d} q}{(2 \pi)^2} V_q^1 V_{-q}^1 \operatorname{Im}\left[D_{11}^r(q, u)\right](1-e^{-i u t}-iut)-i E_\alpha t\right)\\
        &\approx \exp \left(-\int_{\tilde{\Delta}}^{\infty} \mathrm{d} u \frac{1}{u^2} \int_{\frac{c_v}{R_{\Delta}}}^{\frac{1}{\ell_\Delta}} \frac{q \mathrm{~d} q}{(2 \pi)^2} \abs{\frac{2\sqrt{2}i\Delta{\omega_c^*} R_\Delta}{\pi q^2\ell_\Delta^2}}^2 \frac{2\pi k_F q u}{\varphi^2 q^4+k_F^2 u^2}(1-e^{-i u t}-i u t)-iE_\alpha t\right)\\
\end{aligned}
\end{equation}
we now perform the integration over $q$ first and keep the leading order terms in $p$
\begin{equation}
    \begin{aligned}
    F_2(t)&\approx-\abs{\frac{2\sqrt{2}\Delta{\omega_c^*} R_\Delta}{\pi \ell_\Delta^2}}^2\int_{\tilde{\Delta}}^{\infty}\dd u\frac{R_\Delta}{2\pi c_v k_F  u^3}(1-e^{-i u t}-i u t)\\
    &=-\abs{\frac{2\sqrt{2}\Delta{\omega_c^*} R_\Delta}{\pi \ell_\Delta^2}}^2t^2\int_{\tilde{\Delta}t}^{\infty}\dd x\frac{R_\Delta}{2\pi c_v k_F  x^3}(1-e^{-i x }-i x)\\
    &\approx-\frac{t^2{\omega_c^*}^2}{c_v\pi^3(2p+1)}\ln(2p+1),
    \end{aligned}
\end{equation}
%&=-\abs{\frac{2\sqrt{2}\Delta{\omega_c^*} R_\Delta}{\pi \ell_\Delta^2}}^2 \frac{ R_\Delta}{2\pi  c_v k_F} \left(-\frac{1}{2u^2}-\frac{i e^{-i u t}(u t+i)}{2 u^2}+\frac{1}{2} t^2 \operatorname{Ei}(-i u t)+i\frac{t}{u}\right)\Bigg|_{u=\tilde{\Delta}}^\infty\\
%&\approx-\frac{2{\omega_c^*}^2}{c_v\pi^3(2p+1)}\left(\frac{3}{2}-ln(\abs{t}\Tilde{\Delta})-\gamma+sgn(t)\frac{i\pi}{2}\right)t^2+O(\frac{\Tilde{\Delta}}{2p+1})\\
from the second line to the last line, we assumed $\Tilde{\Delta}t\ll1$, and for large $t$, the integral diverges as $t$. We find $F_2(t)$ is roughly a Gaussian. Thus, in the frequency domain, it's also a Gaussian centered at the unperturbed energy with FWHM $\sim\sqrt{\log(p)/p}$.

\subsection{F. Half-filling spectral function: electron perspective}

Now we take the electron perspective. Under linked cluster approximation,

\begin{equation}
    f(u)=-\frac{1}{u^2} \int_0^\infty  \frac{q \mathrm{~d} q}{(2 \pi)^2} V_q V_{-q} \operatorname{Im}\left[\chi^r(q, u)\right],
\end{equation}
where $V_q=2\pi e^2/\epsilon q$ is the Coulomb interaction and $\chi$ is the density response function of the system, which was evaluated to be 
\begin{equation}
    \chi^r(q, u)=\frac{q}{2\pi}\frac{1}{i k_F \omega/q^2-\ell_B E_C}.
\end{equation}
Thus, 
\begin{equation}
    \operatorname{Im}\left[\chi^r(q, u)\right]=-\frac{q^3}{2\pi}\frac{ k_F u}{\ell_B^2 E_C^2 q^4+k_F^2 u^2},
\end{equation}
and
\begin{equation}
\begin{aligned}
        f(u)&=\frac{1}{u^2} \int_0^\infty \frac{q \mathrm{~d} q}{(2 \pi)^2} \abs{\frac{2\pi E_C \ell_B}{q}}^2 \frac{q^3}{2\pi}\frac{ k_F u}{\ell_B^2 E_C^2 q^4+k_F^2 u^2}\\
        &=\frac{1}{u} \int_0^\infty \dd q \frac{E_C^2 \ell_B}{2 \pi} \frac{q^2}{\ell_B^2 E_C^2 q^4+k_F^2 u^2}\\
        &=\frac{E_C^{\frac{1}{2}}}{4\sqrt{2}}u^{-\frac{3}{2}}.
\end{aligned}
\end{equation}
We proceed to evaluate (omit the constant energy shift term at the moment) 
\begin{equation}
    \begin{aligned}
        F_2^1(t)&=-\int_0^{\infty} \mathrm{d} u f(u)\left(1-e^{-i u t}\right)\\
        &=-\frac{E_C^{\frac{1}{2}}}{4\sqrt{2}}\frac{2\left(\sqrt[4]{-1} \sqrt{\pi} \sqrt{t} \sqrt{w} \operatorname{erf}(\sqrt[4]{-1} \sqrt{t} \sqrt{w})+e^{-i t w}-1\right)}{\sqrt{w}}\Bigg|_{w=0}^\infty\\
        &=-\sqrt{i\omega_0 t}
    \end{aligned}
\end{equation}
where $\omega_0=\pi E_C/8$ and we use the asymptotic series expansion of error function $\operatorname{erf}(x)$ for $\abs{x}\rightarrow \infty$.

To get the spectral function, for the particle side, we perform Fourier transformation of $\exp(F_2(t)-i E_e t)$,
\begin{equation}
    \begin{aligned}
        A_+(\omega)&=\int_{-\infty}^\infty e^{-\sqrt{i \omega_0 t}-i E_e t+i\omega t}\dd t\\
        A_+(\omega+E_e)&=-\frac{i \sqrt{\pi} e^{-\frac{1}{4 \tilde{\omega}}} \operatorname{erfi}\left(\frac{-1+2 \sqrt{i t} \tilde{\omega}}{2 \sqrt{\tilde{\omega}}}\right)}{2 \tilde{\omega}^{3 / 2}}-\frac{i e^{i t \tilde{\omega}-\sqrt{i t}}}{\tilde{\omega}} \Bigg|_{t=-\infty}^\infty\\
    \end{aligned}
\end{equation}
where $\tilde\omega=\omega/\omega_0$ and we use the asymptotic form of the imaginary error function $\operatorname{erfi}(x)$ for  $\abs{x}\rightarrow \infty$, we get (insert back the energy shift term, $\bar{\omega}$, to fix the first moment),

\begin{equation}
    A_+(\omega+E_e-\Bar{\omega})=\frac{1}{2}\Theta(\omega)\sqrt{\pi}e^{-\frac{\omega_0}{4\omega}}\left(\frac{\omega_0}{\omega}\right)^\frac{3}{2}.
\end{equation}
Note that the above expression is not correct in UV, and a cutoff $\Omega$ has to be introduced. This is not a surprise since the density response we used for CFL is not UV complete.

\subsection{G. Details about exact diagonalization calculation}
Numerical studies could also contribute to our understanding of the electronic excitations in FQH states, at least for small $p$. We perform exact diagonalization calculations (codes available at \cite{DVN/PJK5G8_2024}, modified based on codes developed by Ryan Mishmash\cite{halled2017}) for up to $10$ electron on a sphere to justify our analysis. We use chord distance as the distance between electrons and the energy is shown in units of the typical coulomb energy scale $E_C$. Since this simulation could not reflect the energy cost of adding/removing an electron from experimental samples, we put in by hand $\mu_\pm=0.5E_C$ to account for this charging energy. The spectral function is obtained by evaluating the eigenstates $\ket{i}^\pm$ and corresponding eigenenergy $E_i^\pm$ in $\mathcal{H}_{N\pm 1}$ and calculating their overlaps $O_i^\pm$ with $c_{QQQ}^\dagger\ket{g}_N$($c_{QQQ}\ket{g}_N$). The spectral function is thus given by $A(\omega)=\sum_i \delta\left(\omega \mp \left(E_i^\pm -E_g^\pm +\mu_\pm\right)\right)O_i^\pm$, where $E_g^\pm$ is the ground state energy in $\mathcal{H}_{N\pm 1}$.

\begin{figure}[!ht]
    \centering
    \hspace{-0.5cm}
    \includegraphics[width=0.35\linewidth]{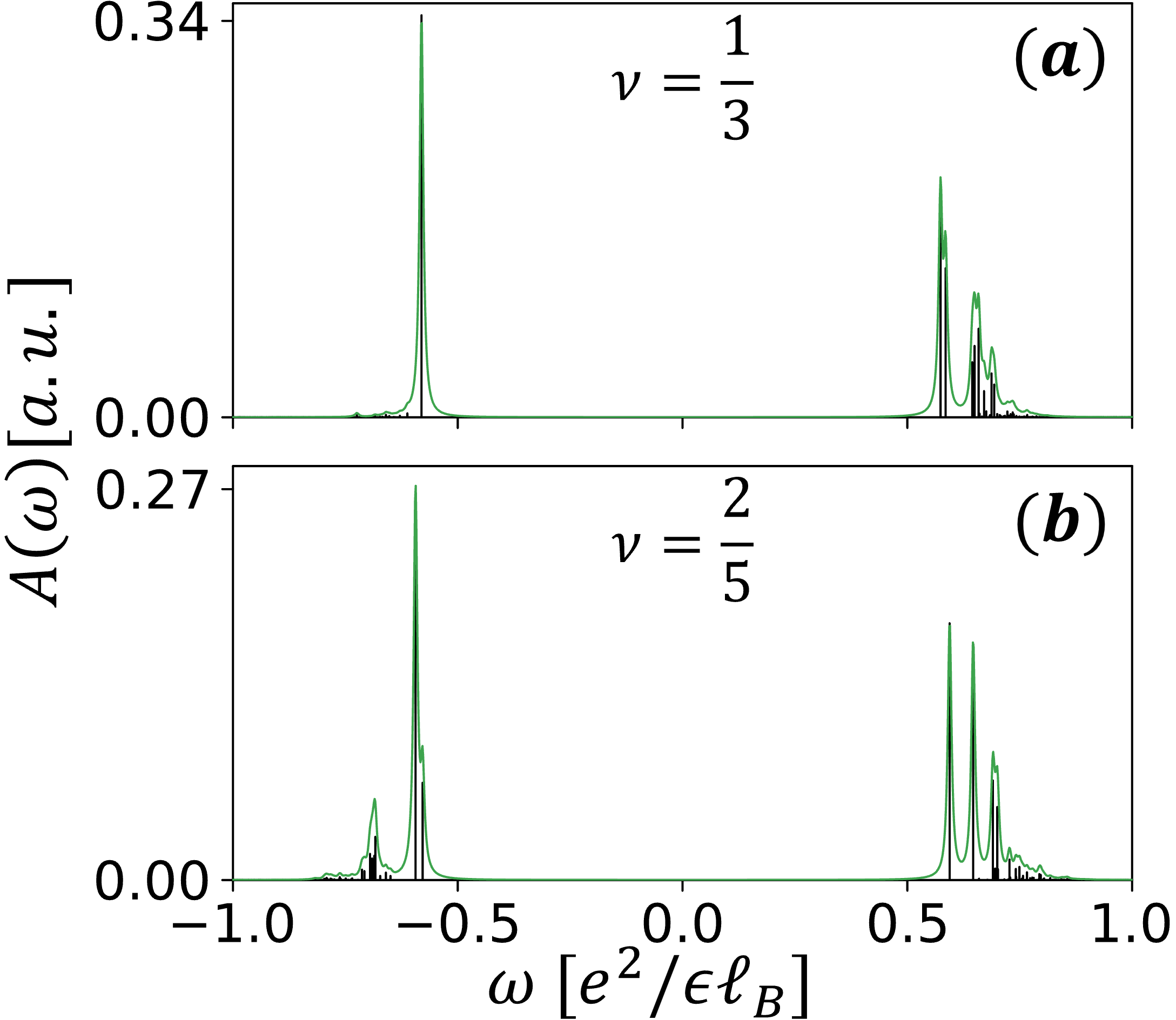}
    \caption{Electronic spectral function obtained from exact diagonalization calculation on a sphere. (a) $N=9$  and $Q=12$, which is the finite size realization of $\nu=1/3$, (b) $N=10$  and $Q=10.5$ which is the finite size realization of $\nu=2/5$. In the calculations, we use chord distance as the distance between electrons. A finite lifetime, $0.005E_C$, and a chemical potential $\mu_\pm=0.5E_C$ have been introduced.}
    \label{edfig}
\end{figure}

Fig.\ref{edfig} shows the spectral function $A(\omega)$ for $v=1/3,2/5$ from exact diagonalization calculations. Qualitatively, we find a series of peaks that become broader and broader with smaller and smaller spectral weights as the corresponding CF states have higher and higher energy. The broadening of high-energy peaks could be understood easily since they have more channels to decay. We remark that the existence of a series of gapped peaks demonstrates CF theory captures the multi-gapped spectrum of this complicated many-electron system. 
%On the other hand, we notice there are only two instead of three peaks in the hole excitation of $\nu=2/5$, which contradicts what we expected from the meanfield CF theory. This missing peak may imply the linear dependence of the corresponding meanfield CF states or simply come from the finite size effect. 
We want to emphasize that the finite size effect is prominent in these exact diagonalization calculations. It is found that for a system with a smaller number of electrons, the number of peaks is smaller, and the peaks at lower energy have more spectral weight compared with a system with more electrons. We believe these finite size effects exaggerate the weight of coherent peaks in the exact diagonalization calculations, especially for cases with larger $p$ (since there's a smaller number of CFs within each $\Lambda$-levels). Therefore, it's hard to obtain the electronic excitation in CFL reliably from exact diagonalization calculations without further refinement.

Another numerical approach was applied to this problem. In recent works\cite {gattu2023stm, pu2023fingerprints}, CF theory is used to obtain the guessed wave functions, which are assumed to capture the low energy Hilbert space well. The electron spectral function (shown in Fig.\ref{half_filling}(c) is obtained by diagonalization within this truncated Hilbert space. Our analysis above for large $p$ is a good compensation for their results at small $p$ due to computational limitations and a comprehensive bridge between the behavior at small and large $p$ limits. Specifically, the electron spectral function at different fillings in their calculation seems to approach our result Eq.\ref{elec_spec} as the filling factor approaches $1/2$.

\twocolumngrid
\bibliography{aps}

\newpage

\end{document}